\documentclass[11pt]{article}

\usepackage[hmargin=1in,vmargin=1in]{geometry}
\usepackage[dvipsnames,table]{xcolor}
\usepackage{amsmath,amssymb,amsfonts,amsthm}
\usepackage{graphicx}
\usepackage{xspace}
\usepackage{framed}
\usepackage{thmtools}
\usepackage{thm-restate}
\usepackage{nicefrac}
\usepackage{mathtools}

\usepackage[hyperfootnotes=false]{hyperref}
\hypersetup{
  colorlinks=true,
  urlcolor=Blue,
  citecolor=red!60!brown,
  linkcolor=BrickRed,
  breaklinks,
  unicode
}
\usepackage[capitalise]{cleveref}

 \usepackage{mathpazo}
\usepackage{setspace}
\newtheorem{theorem}{Theorem}[section]
\newtheorem{corollary}[theorem]{Corollary}
\newtheorem{lemma}[theorem]{Lemma}
\newtheorem{proposition}[theorem]{Proposition}
\newtheorem{definition}[theorem]{Definition}
\newtheorem{remark}[theorem]{Remark}
\newtheorem{hypothesis}[theorem]{Hypothesis}
\newtheorem{claim}[theorem]{Claim}

\newtheorem{fact}[theorem]{Fact}

\newcommand{\eps}{\varepsilon}

\newcommand{\poly}{\mathrm{ poly}}
\newcommand{\cost}{\mathrm{cost}}
\newcommand{\polylog}{\mathrm{ polylog}}
\renewcommand{\tilde}{\widetilde}

\DeclareMathOperator*{\val}{val}

\newcommand{\HH}{\mathcal{H}}

\newcommand{\OPT}{\operatorname{OPT}}
\newcommand{\cov}{\operatorname{cov}}

\newcommand{\E}{\mathbb{E}}
\newcommand{\Var}{\mathrm{Var}}

\renewcommand{\le}{\leqslant}
\renewcommand{\leq}{\leqslant}

\renewcommand{\geq}{\geqslant}

\title{ Almost Optimal Lower Bounds under Almost ETH for\\ Approximating  Euclidean $k$-means}
\author{Karthik C.\ S.\footnote{This work was supported by the National Science Foundation under Grants
CCF-2313372 and CCF-2443697.
}\\ Rutgers University\\\vspace{0.1in}
	\texttt{karthik.cs@rutgers.edu}\and Dor Minzer\footnote{This work was supported by the National Science Foundation under Grants CCF-2227876 and
 CCF-2239160.}\\ MIT\\\vspace{0.1in}
	\texttt{dminzer@mit.edu} }
\date{}
\begin{document}
\maketitle
\begin{abstract}
In the Euclidean $k$-means problem, we are  given a set of $n$ points in $\mathbb{R}^d$, and the goal is to find $k$ representative points in space so as to minimize the sum of the squared distances from each point to its closest representative.
Algorithmically, de la Vega, Karpinski, Kenyon, and Rabani [STOC'03] and Kumar, Sabharwal, and Sen [JACM'10] obtain a $(1+\varepsilon)$-approximation for
high-dimensional  Euclidean $k$-means in time $2^{(k/\varepsilon)^{O(1)}} \cdot dn^{O(1)}$, which was subsequently improved to $2^{\tilde O(k/\varepsilon)} + O(ndk)$ by Feldman, Monemizadeh, and Sohler [SoCG'07]. On the lower-bound side, a recent work of Cohen-Addad, Karthik, Saulpic, and Schwiegelshohn [SoCG'26] proposed the Exponential Time for Expanders Hypothesis (XXH), which asserts exponential-time hardness for $(1+\varepsilon)$-approximating Vertex Cover on vertex expanders, and used it to prove nearly matching running time lower bound whenever $k=\tilde\omega(1/\varepsilon)$.\vspace{0.1cm}

We observe that XXH does not hold in the parameter regime needed for the above result, and show how to repair their reduction under a more standard assumption. 
 More precisely, we show that assuming the Almost Exponential Time Hypothesis for 3-SAT (Almost ETH), for every constant
$\beta\in(0,1)$, Euclidean $k$-means in $\mathbb{R}^d$ has no
$(1+\varepsilon)$-approximation algorithm running in time
$2^{(k/\varepsilon)^{1-\beta}}\operatorname{poly}(n,d)$, whenever $k=\tilde\omega(1/\varepsilon)$.  
By modifying the construction, we also obtain a similar result for approximating the Euclidean $k$-median problem.
\vspace{0.1cm}



\end{abstract}
\thispagestyle{empty}\clearpage
\hypersetup{pageanchor=true}
\pagenumbering{arabic}

\clearpage
\hypersetup{pageanchor=true}
\pagenumbering{arabic}

\section{Introduction}
\label{sec:introduction}

In this paper, we prove conditional time lower bounds under Almost ETH\footnote{The Almost Exponential Time Hypothesis (Almost ETH), introduced in \cite{Manurangsi26}, asserts that, for every constant $\gamma\in(0,1)$, no algorithm decides satisfiability of 3-SAT instances on $n$ variables and $O(n)$ clauses in time $O(2^{n^{1-\gamma}})$. } for Euclidean $k$-means and $k$-median in both the approximation and exact settings.

\subsection{On Approximating Euclidean $k$-means}

In the Euclidean $k$-means problem, we are given a set of $n$ points in
$\mathbb R^d$, and the goal is to choose $k$ centers in $\mathbb R^d$
minimizing the sum of squared distances from each point to its closest
center. This problem has attracted substantial attention over the last half a century across diverse research communities, including operations research, machine learning, and theoretical computer science.

Computationally, the problem is NP-hard even when $d=2$ and $k$ is part of the
input~\cite{megiddo1984complexity,MahajanNV12}, or when $k=2$ and $d$
is part of the input~\cite{dasgupta2009random,AKP24}. On the other hand, the classical exact
algorithm of~\cite{InabaKI94} 
, which runs in time
$O(n^{kd+1})$~\cite{InabaKI94}, remains the state-of-the-art for exact solutions.

While the study of general polynomial-time approximation algorithms for the problem remains an active line of research~\cite{Cohen-AddadS19,Cohen-AddadSL21,Cohen-AddadEMN22,Cohen-AddadSL22,CharikarCGGLW25,CharikarCGGLW26}, there is also a long line of work studying $(1+\varepsilon)$-factor \emph{approximation schemes} for Euclidean
$k$-means whose running time is polynomial in $n$ and $d$ (but may
depend exponentially on $k$
and $1/\varepsilon$). 
The works~\cite{VegaKKR03,KumarSS10}
gave $(1+\varepsilon)$-approximation algorithms running in time
$2^{(k/\varepsilon)^{O(1)}}(nd)^{O(1)}$. Subsequent works~\cite{FeldmanMS07,Jaiswal0S14,jaiswal2015improved,AbbasiBBCGKMSS23,bhore2025coresetstrikesbackimproved} improved the
running time to $2^{\widetilde O(k/\varepsilon)}+O(ndk)$ and
$O(nd\cdot 2^{\widetilde O(k/\varepsilon)})$.
These results leave open whether an exponential dependence on
$k/\varepsilon$ is necessary. In particular, one may ask the following:

\begin{center}
\emph{Is there an $(1+\varepsilon)$-approximation scheme for Euclidean $k$-means }\\ \emph{ on $n$ points in $\mathbb{R}^d$ with
running time $\poly(n,d)\cdot 2^{o(k/\varepsilon)}$?}
\end{center}

Standard hardness reductions already provide conditional time lower bounds 
for $(1+\varepsilon)$-approximating Euclidean $k$-means
in the following two regimes for every constant $\beta\in(0,1)$:
\begin{itemize}
    \item $k=\Theta(n)$ and $\varepsilon=\Theta(1)$:
    Starting from Almost-ETH
    and applying state of the art PCPs \cite{dinur2007pcp}, it is easy to show
    that for some sufficiently small constant
    $\delta>0$, there is no $2^{n^{1-\beta}}$-time algorithm
    that approximates Vertex Cover within a factor
    of $1+\delta$, even on sparse graphs. By a
    standard reduction (see for example \cite{LeeSW17}), this implies that Euclidean
    $k$-means cannot be approximated within a factor
    of $1+\Omega(\delta)$ in time $2^{n^{1-\beta}}$, albeit when
    $k$ is linear in $n$ (the number of input points).

    \item $k=\Theta(1)$ and $\varepsilon=\Theta(1/n)$:
    Starting from Almost ETH, it is easy to show,
    for example by following the reductions
    in~\cite{dasgupta2009random,AKP24}, that for
    some constant $\delta>0$, Euclidean $2$-means
    cannot be approximated within a factor of
    $1+\delta/n$ in time $2^{n^{1-\beta}}$.
\end{itemize}
However, these lower bounds do not rule out, for
example, an algorithm running in time
$2^{O(k+1/\varepsilon)}\cdot\poly(n,d)$.
In particular, the above lower bounds do not directly apply in other settings, such as $k = \Theta(n^{2/3})$, $\eps = \Theta(n^{-1/3})$. 

To address this problem, Cohen-Addad, Karthik, Saulpic, and
Schwiegelshohn~\cite{CKSS26} proposed the Exponential Time for Expanders
Hypothesis (XXH). 
Roughly, it asserts that on regular small-set vertex expanders, no subexponential-time algorithm can distinguish the case where some set of $n/2$ vertices leaves at most a $\zeta$ fraction of the edges uncovered from the case where every such set leaves at least a $\delta$ fraction uncovered, where $\zeta\leq\delta^5/10$.\footnote{This condition is required in their reduction to the Euclidean $k$-means problem.} Assuming XXH, they show that for every $\beta>0$, Euclidean $k$-means on $n$ points in $\mathbb R^d$ has no $(1+\varepsilon)$-approximation algorithm running in time $2^{(k/\varepsilon)^{1-\beta}}\operatorname{poly}(n,d)$ whenever $1/\varepsilon=o(k/\polylog(k))$ (with some additional structural conditions on $k$ and $\varepsilon$).
However, as we show in Appendix~\ref{sec:XXH_false}, in the 
regime where $\zeta\leq\delta^5/10$, the problem can be solved in polynomial time by the Goemans-Williamson algorithm~\cite{GoemansW95}.  Nevertheless, building on their high-level ideas, we establish the following under the more standard Almost ETH assumption.

\begin{theorem}[Informal version of
\Cref{thm:bmv-kmeans-formal}]
\label{thm:intro-bmv}
Assuming Almost ETH, for every constant $\beta\in(0,1)$, Euclidean
$k$-means on $n$ points in $\mathbb R^d$ has no
$(1+\varepsilon)$-approximation algorithm running in time    $2^{(k/\varepsilon)^{1-\beta}}
    \operatorname{poly}(n,d)$ whenever $1/\varepsilon=o(k/\polylog(k))$ and $k/\varepsilon=o(n/\polylog(n))$.
\end{theorem}

This improves over \cite{CKSS26} by establishing the conditional time lower bound under the relatively standard, deterministic Almost ETH assumption. We discuss the technical details of our proof and compare it with \cite{CKSS26} in detail in \Cref{sec:overview}.

\subsection{On Approximating Euclidean $k$-median}
We next consider the Euclidean $k$-median problem. In this problem we are given $n$ points in $\mathbb R^d$ and an integer $k$, and the goal is to find a set of $k$ centers in $\mathbb R^d$ so as to minimize the sum of distances of each input point to its closest center. As with Euclidean $k$-means, there is a long line of work studying $(1+\varepsilon)$-approximation schemes for this problem whose running time is polynomial in $n$ and $d$ but may depend exponentially on $k$ and $1/\varepsilon$. The works~\cite{BadoiuHI02,KumarSS10} gave $(1+\varepsilon)$-approximation algorithms running in time
$2^{(k/\varepsilon)^{O(1)}}\operatorname{poly}(n,d)$, and subsequent works~\cite{FeldmanMS07,FL11,cohen2021improved,cohen2022towards,AbbasiBBCGKMSS23,bhore2025coresetstrikesbackimproved} improved the runtime to $2^{\widetilde O(k/\varepsilon^{2})}\operatorname{poly}(n,d)$. Our next result shows that this is essentially tight up to the exponent of $\eps$.



\begin{theorem}[Informal version of
\Cref{thm:bmv-kmedian-formal}]
\label{thm:k_median}
Assuming Almost ETH, for every constant $\beta\in(0,1)$, Euclidean $k$-median   on $n$ points in $\mathbb R^d$ has no
$(1+\varepsilon)$-approximation running in time    $2^{(k/\varepsilon)^{1-\beta}}
    \operatorname{poly}(n,d)$ whenever $1/\varepsilon=o(k/\polylog(k))$ and $k/\varepsilon=o(n/\polylog(n))$.
\end{theorem}

\subsection{On Exact Algorithms for Euclidean $k$-means and $k$-median}
Finally, \Cref{thm:intro-bmv,thm:k_median} also have implications for exact algorithms, much like those obtained in~\cite{CKSS26}. Both these problems  have natural exact algorithms\footnote{Assuming the 1-median can be found efficiently for each cluster.} with running time $n^{O(kd)}$ \cite{InabaKI94,TirnaucaGBM18}, leading to the following question:

\begin{center}
\emph{Is there an exact algorithm for Euclidean $k$-means or Euclidean $k$-median }\\ \emph{ on $n$ points in $\mathbb{R}^d$ with
running time $n^{o(kd)}$?}
\end{center}

For Euclidean $k$-means,  assuming XXH (which is now disproved in \Cref{sec:XXH_false}), in \cite{CKSS26}, the authors apply standard dimensionality reduction tools to show that\footnote{In \cite{CKSS26}, the authors actually rule out runtimes of $n^{(k\sqrt{d})^{1-\beta}}$ and $n^{d^{1-\beta}}$, but their proof straightforwardly generalizes to the claim written here.} for every constant $\beta>0$ there is no exact algorithm running in time $n^{(k^{1-\alpha}d^{(1+\alpha)/2})^{1-\beta}}$, for all constants $\alpha\in[0,1]$. Using the same transformation as in \cite{CKSS26}, \Cref{thm:intro-bmv,thm:k_median} imply the following:

\begin{corollary}
\label{thm:intro-bmv-exact}
Assuming Almost ETH, for every constant $\beta\in(0,1)$ and all $\alpha\in[0,1]$, Euclidean
$k$-means and Euclidean $k$-median on $n$ points in $\mathbb R^d$ has no
 algorithm running in time    $n^{(k^{1-\alpha}d^{(1+\alpha)/2})^{1-\beta}}
   $  whenever $\omega(k^{2\alpha/(1+\alpha)}\polylog(n))\le d\le o(k^2/\polylog(n))$ and $k^{1-\alpha}d^{(1+\alpha)/2}=o(n/\polylog(n))$. 
\end{corollary}

For Euclidean $k$-means, the proof proceeds by applying a random linear map (which can be derandomized) to the point set from \Cref{thm:intro-bmv} to reduce its dimension, and using the Johnson--Lindenstrauss lemma to relate pairwise distances. We also use random projections for the Euclidean $k$-median problem, however we are not able to relate the cost before and after the projection directly (this is because, unlike in the $k$-means problem, the objective cost cannot be expressed solely in terms of pairwise distances between input points). To remedy that one can use Kirszbraun's extension theorem in the same way as in~\cite{MakarychevMR19} to argue that the objective cost still does not change by much. We omit the (tedious) details for both problems.
It is worth noting that whenever $k^{1-\alpha}d^{(1+\alpha)/2}=\tilde\Omega(n)$ (for example, when $k\sqrt{d}=\tilde\Omega(n)$ or $d=\tilde\Omega(n)$), our lower bound is nearly optimal as it almost matches the naive $d\cdot k^n$-time exhaustive search for Euclidean $k$-means.

\vspace{-1ex}
\paragraph{Statement of AI Usage.}
This paper began with an attempt to prove XXH
under more standard assumptions. ChatGPT 5.6 Sol disproved this hypothesis, and subsequently also disproved a weaker version considered by the authors; see Remark~\ref{rem:weakXXH}. The authors then explored refinements of the soundness analysis of~\cite{CKSS26} that eliminate the graph expansion requirement, and instead use an explicit construction of graph colorings with certain properties. ChatGPT 5.6 worked out all of the details, and then managed to generalize the construction to be based on a complexity assumption regarding hypergraph vertex coverage, yielding Theorem~\ref{thm:hypergraph-to-kmeans}. The authors then attempted to prove the hypergraph vertex coverage assumption, and \cite{minzer2026improved} grew out of failed attempts to prove it for small uniformity. They then realized that the uniformity need not be small, and spoon fed ChatGPT 5.6 until it managed to use a variant of Feige's construction to prove the required result, yielding Theorem~\ref{thm:bmv-hypergraph-source}.
Upon the arrival of ChatGPT 6, the authors asked it to extend the result to Euclidean $k$-median, which the model managed to do in one shot.
\vspace{-1ex}
\paragraph{Supplementary Material.} ChatGPT 6 improved and extended the results in this paper. In particular, this includes removing the condition $1/\varepsilon=o(k/\polylog(k))$ from both \Cref{thm:intro-bmv,thm:k_median}, extending \Cref{thm:intro-bmv,thm:k_median} to the $\ell_1$ metric, and even obtaining an exact lower bound like \Cref{thm:intro-bmv-exact} for $k$-median and $k$-means in the $\ell_1$ metric (even though there are no dimensionality reduction tools for that metric). While we have put some effort to improve the exposition of these results, they are still not in a shape to be included in this paper and therefore we have made them available as supplementary material in \cite{GitHub26}.  
\vspace{-1ex}
\paragraph{Organization of the Paper.}
In Section~\ref{sec:overview}, we provide a proof overview of our results. In Section~\ref{sec:bmv-hypergraph}, we prove our conditional time lower bound for the hypergraph coverage problem, and in Section~\ref{sec:explicit-colorings}, we construct a family of vertex colorings of hypergraphs with certain properties that are useful for proving our clustering lower bounds. Section~\ref{sec:kmeans-lower-bound} gives the reduction to the Euclidean $k$-means problem and proves the corresponding approximation lower bounds. Section~\ref{sec:euclidean-kmedian-lower-bound} does the same but for the Euclidean $k$-median problem.
Finally, Appendix~\ref{sec:XXH_false} disproves XXH in the parameter regime used in~\cite{CKSS26}.

\section{Proof Overview}
\label{sec:overview}
\paragraph{Notations:} Given points $X=\{x_1,\ldots,x_n\}\subseteq\mathbb R^d$, an integer $k\in[n]$, and $z\geq 1$, the Euclidean $(k,z)$-clustering problem   asks for a set of $k$ centers $C\subseteq\mathbb R^d$ minimizing $ \sum_{i=1}^n\min_{c\in C}\|x_i-c\|_2^z$, and we denote the minimum cost by $\OPT^{z}_k(X)$.  When $z=1$, this is the $k$-median problem and when $z=2$, this is the $k$-means problem. 

\subsection{Proof Overview of \cite{CKSS26}}

We first describe the proof framework of \cite{CKSS26}, which is the basis of the current paper. Let $G=([n],E)$ be a $d$-regular graph (where $m:=|E|=O(n)$), thought of as an instance of the vertex cover problem that is hard to approximate to a $1+\delta$ factor (for some constant $\delta>0$). We denote by $\alpha_G$ the size of an optimal vertex cover of $G$, and we  assume that $\alpha_G=\Omega(n)$. Suppose we would like to reduce $G$ to a $k$-means instance where $k=o(m)$ and $\varepsilon=\Omega(\delta k/\alpha_G)$. The first step in \cite{CKSS26} is to uniformly and independently color each vertex $v$ in $G$ with a color in $c_v\in [k]$. For simplicity of exposition, assume that no edge is monochromatic, i.e., that $c_u\neq c_v$ for every $\{u,v\}\in E$. Thus, each edge gets exactly two colors, and the Euclidean $k$-means instance is obtained by mapping an edge $\{u,v\}\in E$ to the point $\mathbf e_u+\mathbf e_v+\mathbf e_{n+c_u}+\mathbf e_{n+c_v}\in\mathbb R^{n+k}$, where $\mathbf e_i$ is the standard $i$th basis vector.
Throughout this section, we refer to the first $n$ coordinates as the vertex coordinates and the last $k$ coordinates as the color coordinates. We also refer interchangeably to an edge and its corresponding point whenever the meaning is clear from context.

In the completeness case, let $S\subseteq[n]$ be an optimal vertex cover which we partition as $S_1\dot\cup\cdots\dot\cup S_k$ according to the vertex colors. Because the coloring was chosen uniformly and randomly, with high probability we have that $|S_i|= (1\pm o(1))\alpha_G/k$ for every $i\in[k]$. Assign each edge to one of its endpoints in $S$, breaking ties arbitrarily, and let $C_i$ consist of the points corresponding to edges assigned to vertices in $S_i$. Place the center of each nonempty cluster $C_i$ at $\mathbf e_{n+i} +|S_i|^{-1}\sum_{v\in S_i}\mathbf e_v$. Since no edge is monochromatic, every edge assigned to $S_i$ has exactly one endpoint in $S_i$ so exactly one endpoint of color $i$. Its squared distance from this center is therefore at most $3-1/|S_i|=3-(1\pm o(1))k/\alpha_G$: the vertex coordinates contribute $2-1/|S_i|$, while the color coordinates contribute $1$. Summing over all $m$ points gives a clustering of total cost at most $3m-(1-o(1))km/\alpha_G$.

In the soundness case, we would like to show that a sufficiently cheap clustering yields a small set of vertices covering most edges, but there is one concern that needs to be addressed:  one can obtain cheap clustering cost by exploiting the graph topology without having a small vertex cover. To illustrate this, consider two clusters $C_1$ and $C_2$, each containing $2d$ points. The edges corresponding to $C_1$ form two disjoint stars with $d$ edges each, while those corresponding to $C_2$ form a copy of $K_{2,d}$. Let the two star centers and the two vertices on the small side of $K_{2,d}$ have the same color, and let the set of colors (including multiplicity) of the leaves of each star and  the vertices on the larger side of $K_{2,d}$ be the same. 
The color coordinates then contribute equally to the costs of the two clusters. The vertex coordinates however contribute less in $C_2$: each of the $2d$ vertices which are leaves in $C_1$ belongs to only one edge, whereas each of the $d$ shared neighbors in $C_2$ belongs to two edges. Thus the vertex-coordinates contribution to the  clustering cost is $3d-1$ for $C_1$ and $3d-2$ for $C_2$. Consequently, $\OPT^2_1(C_2)=\OPT^2_1(C_1)-1$, even though both edge sets can be covered by two vertices. This illustrates why comparing with the completeness case clustering is not enough: clusters whose edges share neighbors can cost less than clusters formed from disjoint stars.

To avoid such configurations, \cite{CKSS26} require the input graphs
to have near-perfect small-set vertex expansion by working under  XXH. Recall that
XXH asserts that for suitable positive constants $\delta$ and $\zeta$
with $\zeta\leq\delta^5/10$, and for every constant $\beta>0$, no
randomized algorithm running in time $2^{n^{1-\beta}}$ that takes a $d$-regular graph $G=([n],E)$ with near-perfect small-set vertex expansion as input can distinguish between the following cases with probability at least $0.9$:
\begin{description}
\item[Completeness:] There is a set of $n/2$ vertices covering at least a $(1-\zeta)$ fraction of $E$.
\item[Soundness:] Every set of $n/2$ vertices leaves at least a $\delta$ fraction of $E$ uncovered.
\end{description}

From the completeness calculation above, allowing for a small fraction of edges to
remain uncovered,  \cite{CKSS26} obtain
a clustering of cost at most $3m-(1-7\zeta)kd$.  On the other hand, their soundness analysis, which we explain below, rules out a clustering of cost at most $3m-(1-\delta^5)kd$.  

In the soundness analysis, for each cluster $C_i$, they consider  the underlying subgraph induced by the edges forming the cluster. They first express the contribution of the vertex coordinates through degrees in the underlying subgraph: for a cluster $C_i$, let $m_i:=|C_i|$, and let $d_{i,v}$ be the number of edges from $C_i$ incident to vertex $v$. The centroid of $C_i$ has value $d_{i,v}/m_i$ in coordinate $v$, so the vertex coordinates contribute exactly $2m_i-m_i^{-1}\sum_{v\in[n]}d_{i,v}^2 
$ to the cluster cost. The fact that the clustering is cheap amounts to 
$\sum_{v\in[n]}d_{i,v}^2\geq (1-o(1))dm_i$ for a typical $i$. 

The next step in their argument is to show that this is almost an equality.
One always has the trivial upper bound 
$\sum_{v\in[n]}d_{i,v}^2\leq \max_{v}d_{i,v}\sum_{v\in[n]}d_{i,v}\leq 2dm_i$, and vertex expansion is used to establish the sharper upper bound of $(1+o(1))dm_i$. Indeed, letting $G_i$ be the subgraph induced by $C_i$, we know that $|V(G_i)|\leq 2m_i = o(n)$, and 
if this sum was to be at least $(1+\delta)dm_i$, then by averaging there is a set $S$ of fractional size at least $\frac{\delta}{2}$ such that each $v\in S$ has $d_{i,v}\geq \frac{\delta}{2} d$. In that case $V(G_i)$ contains at least $\frac{\delta^2}{8} d|V(G_i)|$ edges (namely, the edges counted by $d_{i,v}$ for $v\in S$), contradicting the small-set expansion. 

Overall, 
$\sum_{v\in[n]}d_{i,v}^2 = (1\pm o(1))d m_i$, which implies that almost all of the contribution to the left hand side comes from vertices $v$ with $d_{i,v}\geq (1-o(1))d$, and the rest of this argument shows that taking all such $v$'s gives a small vertex-cover.



\subsection{Overcoming the Need for Vertex Expansion}
Our approach starts by analyzing the same embedding as above, but we consider the effects of the vertex coordinates and the color coordinates together. To explain this, we denote $D_i:=m_i^{-1}\sum_{v\in[n]}d_{i,v}^2$, 
and recall that the vertex coordinates contribute $2m_i-D_i$ to the cost. 
To better understand the contributions from the color coordinates we denote by $\gamma_i$ the fraction of unordered pairs of edges in $C_i$ whose
color sets are disjoint, and by $\kappa_i$ the fraction  whose color sets are identical. A quick calculation shows that the $k$-means cost of $i^{\text{th}}$ cluster is
$\OPT^2_1(C_i)=3m_i-1-z_i$ where $z_i:=D_i-(m_i-1)\gamma_i+(m_i-1)\kappa_i$, and our goal is therefore to upper bound $z_i$.
\vspace{-1ex}
\paragraph{Coloring Properties.}
To bypass the need for expansion we use additional properties of the random coloring: 
\begin{enumerate}
    \item Each color class contains $(1+o(1))n/k$ vertices and only $O(m/k)$ edges have both endpoints of the same color.
    \item Every vertex has at most three neighbors of any fixed color.
    \item Any pair of distinct colors occurs together on at most $O(\log k)$ edges.
\end{enumerate}
Using the fact that the number of colors we used is significantly larger than the degree, it is easy to show that a uniformly random coloring of the vertices satisfies all three properties.

To see the relevance of these properties to the soundness analysis, note that the third property immediately implies that $m_i\kappa_i = O(\log k)$, which is $o(d)$ in our case. Thus, understanding the saving $z_i$ reduces to understanding the tradeoff between the parameters $D_i$ and $\gamma_i$.

\vspace{-1ex}
\paragraph{Identifying a Dominant Color.}  
Let $c_i$ be the most frequent color in cluster $i$, let $h_i$ be the number of its edges containing $c_i$, and let $R_i:=m_i-h_i$ be the number not containing it.
We show that $\sum_{i\in[k]}R_i = o(m)$, which says that in most clusters, most edges contain the dominant color. 
We henceforth discard the edges accounted by $R_i$, and assume for simplicity that all of the quantities we studied remain unchanged (the more general case follows similarly, by accounting how much the deletion of $R_i$ changes). Our next goal is to study the structure of the clusters, and show that they must behave like in the completeness case.

\vspace{-1ex}

\paragraph{Bounding $z_i$.}   
By definition, as $R_i=0$ we get that $\gamma_i=0$ so  $z_i=D_i+o(d)$, and we explain the connection between the value of $z_i$ and the way the cluster $C_i$ looks like. A crude upper bound on $D_i$ as before gives that $z_i\le (2+o(1))d$, and by inspection a value close to $2d$ is only attained when the subgraph induced by the edges of $C_i$ forms (roughly) a $d$ regular graph (most $d_{i,v}$ are close to $d$). In that case, the subgraph induced by $C_i$ is a (1) bipartite graph, (2) its left side is vertices colored $c_i$, and its right side is vertices colored $c\neq c_i$, (3) it is (basically) $d$-regular. In particular, most vertices on the left hand side have $d\gg 3$ neighbors colored $c_i$, violating the second property of the coloring. 
This gives that $z_i\leq (2-\Omega(1))d$, and pushing this argument further actually gives that $z_i\leq (1+o(1))d$. As the total cost of the clusters is $3m-(1-o(1))kd$, we get that
$\sum_{i\in[k]}z_i\geq(1-o(1))kd$, and hence $z_i = (1-o(1))d$ for almost all $i$, so intuitively the clusters $C_i$ should look like unions of $d$-stars.

\vspace{-1ex}

\paragraph{Extracting a Small Vertex Cover.}
Our next goal is to extract a small vertex cover, and we first assign an edge in $C_i$ to its unique endpoint colored $c_i$. Let
$A_v:=\sum_{i:c_v=c_i}d_{i,v}$ be the total number of edges assigned
to vertex $v$ across all clusters, and think of $x_v:=A_v/d$ as a
fractional selection of vertices. Since every edge is assigned at
most once, we have $\sum_v A_v\leq m$, so the total weight of this
fractional selection is at most $m/d=n/2$. We next show that these
weights account for almost all edges. Note that,
\[
\sum_v A_v^2
\geq \sum_v\sum_{i:c_v=c_i}d_{i,v}^2=\sum_i\sum_{v:c_v=c_i}d_{i,v}^2=\sum_i m_i\left(D_i-
\frac{1}{m_i}\sum_{v:c_v\neq c_i}d_{i,v}^2\right).
\]
The second term is negligible since the second coloring property
ensures that $d_{i,v}\leq3$ for every vertex not colored $c_i$.
Moreover, the contribution of $\kappa_i$ is negligible, so
$\sum_v A_v^2\geq\sum_i m_i z_i-o(md)$. Using the information about each individual $z_i$ is useful we know that essentially for all $i$, $z_i = (1-o(1))d$, giving (some care is needed to ensure the $m_i$ are mildly balanced) 
\[
\sum_i m_i z_i \geq (1-o(1))d\sum\limits_{i}m_i = (1-o(1))dm,
\]so overall  
$\sum_{v}A_v^2\geq (1-o(1))dm$. Rewriting this using $x_v$ gives
$\sum_v x_vA_v\geq(1-o(1))m$,
implying $\{x_v\}_v$ is a fractional solution of size at most $n/2$ and cover at least $(1-o(1))m$. Choosing the
$n/2$ vertices of largest $x_v$ thus gives a set of $n/2$
vertices covering $(1-o(1))m$ edges.

\subsection{Moving to Hypergraph Vertex Coverage}

Having removed the need for vertex expansion, it remains to obtain an almost-linear-size reduction from $3$-SAT to Vertex Cover instances with the coverage guarantees required by our new analysis. Towards this end, note that naive reductions, such as padding a given vertex-cover instances by $K_{d,d}$, do not work as the gap between completeness and soundness becomes too weak. Next, we note that existing hardness results obtained through the PCP theorem (namely, the ones following from the FGLSS reduction or from known dictatorship tests) do not have the correct cover in the completeness size. A result in the literature that has the properties that we need are the UGC-based hardness results, such as~\cite{AKS11}, but the size of the reduction is a large polynomial (and this is necessary~\cite{AroraBS10}).

\vspace{-1ex}
\paragraph{Hypergraph Vertex Coverage.} To circumvent this issue we observe that the $k$-means reduction extends naturally from vertex cover in graphs to coverage in (constant uniformity) hypergraphs, as defined next. 

\begin{definition}
Let $\HH=(V,\mathcal E)$ be a hypergraph and let $k\in\mathbb N$.
For $S\subseteq V$, define
$\cov_{\HH}(S):=|\{e\in\mathcal E:e\cap S\neq\varnothing\}|/|\mathcal E|$.
The maximum fraction of hyperedges covered by at most $k$ vertices is
$\OPT_{\HH}(k):=\max_{S\subseteq V,\ |S|\leq k}\cov_{\HH}(S)$.
\end{definition}

A hypergraph is $t$-uniform if every hyperedge contains exactly $t$ vertices, and it is $\Delta$-regular if every vertex belongs to exactly $\Delta$ hyperedges. The codegree of vertices $u,v$ is the number of hyperedges $e$ containing both of them, and we let $\Lambda(\HH)$ for the maximum codegree of $\HH$ among all pairs $u\neq v$. We prove the following conditional time lower bound for the gap hypergraph vertex coverage problem.

\begin{theorem}
\label{thm:bmv-hypergraph-source}
Assuming Almost ETH, for every $\eps>0$, there is a constant $t\geq2$ such that
for every constant $\beta\in(0,1)$, no algorithm running in time
$2^{O(N^{1-\beta})}$ on input a $t$-uniform hypergraph $\HH$ on $N$ vertices, and $k=N/t$, can distinguish between the following cases.
\begin{description}
\item[Completeness:] There is a set of exactly $k$ vertices
that meets every hyperedge in exactly one vertex.
\item[Soundness:] $\OPT_{\HH}(k)\leq1-1/e+\eps$.
\end{description}
In both cases, $\HH$ is regular with degree
$\Delta=\Theta_{\eps}((\log N)^{3(t-1)})$ and maximum codegree at most
$\Delta/(\log N)^2$.
\end{theorem}
The proof of~\Cref{thm:bmv-hypergraph-source} proceeds via a combination of known techniques, and we refer the reader to~\Cref{sec:bmv-hypergraph} for an outline.
\vspace{-1ex}
\paragraph{On the Codegree.} In a simple graph, a given pair of vertices can occur only at a single edge, but the same is not true for hypergraphs. This presents an issue in adapting our analysis from graphs to hypergraphs. Towards this end, it turns out to be sufficient that the codegree of the hypergraph is sufficiently small compared to the degree of a hypergraph, and this is the reason~\Cref{thm:bmv-hypergraph-source} discusses the codegree of $\HH$.

\vspace{-1ex}

\paragraph{From Hypergraph Vertex Coverage to $k$-means.}
We now explain how to modify the reduction to hypergraphs. To avoid overloading the notation $k$ from the definition above, let $K$ denote the desired number of clusters. We color the vertices of a $t$-uniform hypergraph $\HH=(V,\mathcal E)$ with a coloring $\rho\colon V\to[K]$ (which should be thought as a random coloring\footnote{Our actual reduction enumerates an explicit family of $2^{O((\log N)^2)}$ colorings (see \Cref{sec:explicit-colorings}). At least one coloring in this family admits the properties needed for the completeness analysis of the clustering cost, whereas every coloring satisfies the properties needed for the soundness analysis.
}) and delete the hyperedges containing two vertices of the same color. For each remaining hyperedge $e$, we introduce the point:
\begin{equation*}
p_e:=\sum_{v\in e}\mathbf e_v+
\sum_{v\in e}\mathbf e_{N+\rho(v)}
\in\mathbb R^{N+K}.
\end{equation*}
Thus, every point has $t$ vertex coordinates and $t$ color coordinates equal to one. 

In the case the hypergraph $\HH$ was linear, namely the co-degree was $1$, the analysis from before works in basically the same way (because any two hyperedges are either vertex disjoint or share exactly one vertex). In our actual case the codegree of $\HH$ is not $1$, but it is guaranteed to be much smaller than the degree of $\HH$. The analysis in this case relies on the fact that while this may reduce the cluster cost in the soundness case, this change is only negligible. Indeed, as we have seen earlier, the parameter $z_i$ expressing the saving in a cluster is typically relating to the degree of the graph/hypergraph, and the change introduced by codegree will be lower order.

\subsection{Extension to Euclidean $k$-median}
To extend our $k$-means result to $k$-median we use the linearization trick, that has appeared multiple times in fine-grained complexity  (see for example, \cite{Williams18,Chen20,BringmannNusser21,BlankEtAl26}) and even in clustering \cite{Shenmaier21}. The idea is to introduce a new private coordinate $i_p$ for each point $p$ in the $k$-means point set. Then, taking a sufficiently large parameter $T$, we replace the point $p$ with the point $p'=(p,T\cdot e_{i_p})\in\mathbb{R}^{n+k+m}$.
Letting $c_i' = (c_i,u_i)$ be the median of cluster $i$, we show that the centers $(c_i,0)$ that do not use the private coordinate have almost the same cost.
Now, the median cost of the clustering with centers $(c_i,0)$ can be related to the $k$-means cost of the centers $c_i$, thanks to the fact that by Taylor's approximation,
\[
\|(c_i,0)-(p,T\cdot e_{i_p})\|_2
=\sqrt{T^2+\|c_i-p\|_2^2}
= T + \frac{1}{2T}\|c_i-p\|_2^2 + O\left(\frac{\|c_i-p\|_2^4}{T^3}\right).
\]

\section{Lower Bounds for Hypergraph Vertex Coverage: Proof of~\Cref{thm:bmv-hypergraph-source}}
\label{sec:bmv-hypergraph}

The goal of this section is to prove~\Cref{thm:bmv-hypergraph-source}, and the construction proceeds as follows:
\begin{enumerate}
    \item \textbf{The starting point:} Starting with the size efficient PCP construction of~\cite{BMV24}, we apply elementary manipulations to get a corresponding result for label cover instances in which each constraint is a $d$-to-$1$ constraint, where $d$ depends on soundness of the construction. 
    \item \textbf{Regularization:} Applying tools from~\cite{MoshkovitzRaz,DH13}, we further ensure that label cover instance is over a bi-regular graph, where the right degree is a power of $2$.
    \item \textbf{Total disagreement:} The next step is to switch a notion of soundness which is better suited for coverage problems, and the one we use, called agreement soundness, is from~\cite{Mos15}. 
    In particular, we use the transformation from~\cite{Mos15}, which shows how to convert a label-cover instance with small soundness and right degree which is a power of $2$, to a label cover instance with strong agreement soundness.
    \item \textbf{Transferring to a cover problem:} We use the transformation of~\cite{Feige98} to transform our label cover problem into a hypergraph cover problem. This transformation gets a statement in the spirit of~\Cref{thm:bmv-hypergraph-source} with constant degree. To get the gap between the degree and the codegree we blow-up each vertex into $\Theta(\log^3 n)$ many copies. Here, we use an interesting fact from optimization~\cite{CalinescuChekuriPalVondrak2011} regarding submodular optimization, which can be used to show that the blow-up of a hypergraph changes hypergraph coverage by the blow-up factor.
\end{enumerate}

\subsection{The Starting Problem}
We will use the label cover problem in our reductions, defined as follows.
\begin{definition}
An instance of label cover $\Psi$ consists of a bipartite graph $G = (L\cup R, E)$, alphabets $\Sigma_L$, $\Sigma_R$ and a collection of projection constraints $\{\phi_e\colon \Sigma_L\to\Sigma_R~|~e\in E\}$. A constraint $\phi_e$ is called $d$-to-$1$ if for all $\tau\in \Sigma_R$, we have that $|\phi_{e}^{-1}(\tau)|=d$.
\end{definition}
The size of $\Psi$ is defined as $|L|+|R|+|E|$, the alphabet size of $\Psi$ is defined as $|\Sigma_L|+|\Sigma_R|$, and we next define the value of $\Psi$. Fix a label cover instance $\Psi$, and let $e = (u,v)\in E$ be an edge in it. We say that labeling $u$ by $\sigma\in \Sigma_L$ and $v$ by $\tau\in \Sigma_R$ satisfies the constraint on $e$ if $\phi_{e}(\sigma) = \tau$. For labelings $A_{L}\colon L\to\Sigma_L$ and $A_R\colon R\to\Sigma_R$, we define 
\[
\mathsf{val}(A_L,A_R) = \frac{|\{e=(u,v)\in E~|~\phi_e(A_L(u))=A_R(v)\}|}{|E|},
\]
and define the value of $\Psi$ as $\mathsf{val}(\Psi) = \max_{\substack{A_L\colon L\to\Sigma_L\\A_R\colon R\to\Sigma_R}}\mathsf{val}(A_L,A_R)$.
We use the following result from~\cite{BMV24}. We remark that strictly speaking, their result is concerned with weighted label-cover instances, however using the result of~\cite{crescenzi2001weighted} such instances can be transferred to unweighted instances without changing the parameters by much.

\begin{theorem}
\label{thm:bmv-published}
For every $\eps>0$, there is $C>0$ and a polynomial-time reduction that given $3$-SAT instance $\varphi$ produces an instance $\Psi$ of label cover of size at most $n (\log n)^C$ and alphabet size at most $C$, such that the following holds.
\begin{description}
    \item[Completeness:] If $\varphi$ is satisfiable, then $\val(\Psi)=1$.
    \item[Soundness:] If $\varphi$ is unsatisfiable, then $\val(\Psi)\le\eps$.
\end{description}
\end{theorem}

We will next want to establish a version of~\cref{thm:bmv-published} wherein each constraint is $d$-to-$1$ for some $d = d(\eps)$, and we now show an elementary transformation that achieves it.

Fix a label cover instance $\Psi$ as in~\Cref{thm:bmv-published}, and fix $d$ so that $|\phi_e^{-1}(\tau)|\leq d$ for all $e\in E$ and $\tau\in \Sigma_R$. For each edge $e$ define the deficit map $\delta_e\colon \Sigma_R\to \mathbb{N}$ by 
$\delta_e(\tau) = d- |\phi_e^{-1}(\tau)|$ and note that $\sum\limits_{\tau\in\Sigma_R}\delta_e(\tau) = d|\Sigma_R|-|\Sigma_L|$
is independent of $e$. Let $\mathcal{R}$ be an arbitrary set of that size, and for each edge $e\in E$ choose a map 
$h_e\colon \mathcal{R}\to \Sigma_R$ such that $|h_e^{-1}(\tau)|=\delta_e(\tau)$ for every $\tau\in\Sigma_R$. We now define the new label cover instance $\Psi'$, which depends on a parameter $T$ that we will choose in the end. It has the same graph as $\Psi$ but each edge is repeated $T$ times, and it has the new alphabets
\[
\Sigma_R' = \Sigma_R\times \{0,1,\ldots,T-1\},
\qquad 
\Sigma_L' = (\Sigma_L\cup \mathcal{R})\times \{0,1,\ldots,T-1\}.
\]
As per the constraints, for each edge $e\in E$ and  $t\in \{0,\ldots,T-1\}$, we define the constraint on the $t^{\text{th}}$  copy of $e$ by
$\phi_{e,t}'(\sigma,i) = (\phi_e(\sigma),i)$ if $\sigma\in \Sigma_L$, and $\phi_{e,t}'(\sigma,i) = (h_e(\sigma),i+t\pmod{T})$ if $\sigma\in\mathcal{R}$.

The completeness of the reduction is given by the following claim.
\begin{claim}\label{claim:d_to_1_complete}
    If $\val(\Psi)=1$, then $\val(\Psi')=1$.
\end{claim}
\begin{proof}
    Given assignments $A_L,A_R$ that satisfy all of the constraints in $\Psi$, we define the assignments $A_L',A_R'$ to $\Psi'$ by 
    $A_L'(u) = (A_L(u),0)$ and $A_R'(v) = (A_R(v),0)$. It is straightforward to see that they satisfy all of the constraints in $\Psi'$.
\end{proof}

The soundness of the reduction is given in the following claim.
\begin{claim}\label{claim:sound}
    If $\val(\Psi)\leq \eps$, then $\val(\Psi')\leq \eps+\frac{1}{T}$.
\end{claim}
\begin{proof}
    Suppose $A_L',A_R'$ are assignments to $\Psi'$, and fix an edge $e=(u,v)\in E$ in $\Psi$. Let $A_L'(u) = (\sigma,i)$ and $A_R'(v) = (\tau,j)$. If $\sigma\not\in \Sigma_L$, then the constraint of the $t^{\text{th}}$  copy of the edge is satisfied only if $\tau = h_e(\sigma)$ and $j=i+t$. In particular, it must be the case that $j=i+t$, so at most one of the $T$ constraints is satisfied. If $\sigma\in \Sigma_L$, then the constraint is satisfied only if $\tau = \phi_e(\sigma)$. Overall, we conclude that $A_L',A_R'$ satisfy at most $\val(\Psi) + \frac{1}{T}\leq \eps+\frac{1}{T}$ fraction of the constraints.
\end{proof}

Now we have that each constraint in $\Psi'$ is $d$-to-$1$, but the underlying graph is not simple. This can again be fixed by using the technique of~\cite{crescenzi2001weighted}. In summary, we conclude the following result:

\begin{theorem}
\label{thm:bmv-exact}
For every constant $\eps>0$, there are $d>0$, $C>0$ and a polynomial-time reduction that given $3$-SAT instance $\varphi$ produces an instance $\Psi$ of label cover of size at most $n (\log n)^C$, alphabet size at most $C$, and all constraints being $d$-to-$1$, such that the following holds.
\begin{description}
    \item[Completeness:] If $\varphi$ is satisfiable, then $\val(\Psi)=1$.
    \item[Soundness:] If $\varphi$ is unsatisfiable, then $\val(\Psi)\le\eps$.
\end{description}
\end{theorem}

\subsection{Regularization}
We next turn the result from~\Cref{thm:bmv-exact} to a similar result where the underlying label cover graph is bi-regular, and for that we use techniques from~\cite{MoshkovitzRaz,DH13}. More specifically, we use~\cite[Theorem 5.1]{DH13} stated below:
\begin{theorem}\label{thm:apply_dh}
    For all $\delta>0$ there exists a polynomial time reduction mapping a label cover instance $\Psi = ((L\cup R), E, \Sigma_L,\Sigma_R,\{\phi_e\}_{e\in E})$ to a label cover instance $\Psi' = ((L\cup R'),E',\Sigma_L,\Sigma_R,\{\phi_e\}_{e\in E'})$ such that the following holds.
    \begin{enumerate}
    \item \textbf{Completeness:} If $\val(\Psi)=1$, then $\val(\Psi')=1$.
    
    \item \textbf{Soundness:} If $\val(\Psi)\leq \eps$, then $\val(\Psi')\leq \eps+4\sqrt{\delta}$.
    
    \item \textbf{Size and Right Degree:} The graph $(L\cup R',E')$ is right regular, its right degree is at most $\frac{4}{\delta^2}$, and $|R'|=|E|$.
    
    \item \textbf{Left Degree:} If $(L\cup R,E)$ was left regular with left degree $D$, then $(L\cup R',E')$ is left regular with left degree
    \[
        D' \leq \frac{4}{\delta^2}D.
    \]
    
    \item \textbf{Constraints:} If each constraint in $\Psi$ is $d$-to-$1$, then each constraint in $\Psi'$ is $d$-to-$1$.
\end{enumerate}
\end{theorem}
\begin{proof}
    The transformation is the same as in~\cite[Theorem 5.1]{DH13}, and the first three items are stated there. The fourth item follows by inspection of the proof therein, and the last item follows as for each $e'\in E'$, the constraint $\phi_{e'}'$ appears as the constraint $\phi_e$ for some $e\in E$, so it is $d$-to-$1$.
\end{proof}

We now state the version of~\Cref{thm:bmv-exact} for bi-regular graphs:
\begin{theorem}
\label{thm:bmv-regular}
For every $\eps>0$, there are $d>0$, $C>0$ and a polynomial-time reduction that given $3$-SAT instance $\varphi$ produces an instance $\Psi$ of label cover of size at most $n (\log n)^C$, alphabet size at most $C$, all constraints being $d$-to-$1$, and bi-regular with degrees $O_{\eps}(1)$ such that the following holds.
\begin{enumerate}
    \item \textbf{Completeness:} If $\varphi$ is satisfiable, then $\val(\Psi)=1$.

    \item \textbf{Soundness:} If $\varphi$ is unsatisfiable, then $\val(\Psi)\le\eps$.
\end{enumerate}
\end{theorem}
\begin{proof}
   Let $\Psi = ((L\cup R), E,\Sigma_L,\Sigma_R,\{\phi_e\}_{e\in E})$ be the output of the reduction from~\Cref{thm:bmv-exact} for $\eps/3$, and set $
      \delta=\left(\frac{\eps}{12}\right)^2$.
   Applying~\Cref{thm:apply_dh}, we get a label cover instance $\Psi'=((L\cup R'), E',\Sigma_L,\Sigma_R,\{\phi_e'\}_{e\in E'})$ which is right regular, has right degree $
      D\leq \frac{4}{\delta^2}=4\cdot\left(\frac{12}{\eps}\right)^4$,
   has $|R'| = |E|$, and satisfies the completeness and soundness therein. In particular, if $\varphi$ is unsatisfiable, then
      $\val(\Psi')\leq \frac{\eps}{3}+4\sqrt{\delta}
      =\frac{2\eps}{3}$.

   We now switch the sides as in~\cite{MoshkovitzRaz,DH13}. More specifically, we define the label cover instance $\Psi'' = ((R'\cup L), E'', \Sigma', \Sigma_L, \{\phi_{e}\}_{e\in E''})$ as follows: its left side is $R'$, its right side is $L$, and the edge set $E''$ is the same as $E'$ but in reverse, and its right alphabet is $\Sigma_L$. The left alphabet $\Sigma'\subseteq \Sigma_L^{D}$ is thought of as follows: fix $u\in R'$ and let $v_1,\ldots,v_D$ be its neighbours in $\Psi'$. We think of a label to $u$ as a sequence of labels $(\sigma_1,\ldots,\sigma_D)\in \Sigma_L^{D}$ to its neighbours, and we only allow sequences such that there exists $\tau\in \Sigma_R$ such that
   \[
      \phi_{(u,v_i)}(\sigma_i) = \tau\qquad\qquad\forall 
      i=1,\ldots,D.
   \]
   Because each constraint in $\Psi'$ is $d$-to-$1$, it is easy to see that there there $|\Sigma_R| d^{D}$ many such sequences, so that $|\Sigma'| = |\Sigma_R| d^D$. As per the constraints, thinking of the label to $u\in R'$ as $(\sigma_1,\ldots,\sigma_D)$, and of the label $\sigma$ to $v_i$, the $i^{\text{th}}$  neighbour of $u$, the constraint on the edge $(u,v)$ requires that $\sigma_i=\sigma$.
   
   It is clear that $\Psi''$ is a label cover instance, that each constraint in it is $d^{D-1}$-to-$1$ and that $\Psi''$ is left regular with left degree equal to $D$. It is also clear that the transformation preserves perfect completeness and a quick argument shows that it does not increase soundness. Applying~\Cref{thm:apply_dh} to $\Psi''$, with the same value of $\delta$, gives an instance which is right regular and, by the left-degree guarantee in~\Cref{thm:apply_dh}, also left regular. Moreover, in the soundness case its value is at most
$      \frac{2\eps}{3}+4\sqrt{\delta}=\eps$.
   This gives the label cover instance as required.
\end{proof}

For technical reasons, we will need the right degree of our graph to be a prime power. This is easy to achieve by invoking the result of~\cite{DH13} once again, but to do so we need to explain a little bit about how it works.  Given a label cover instance $\Psi$ with right degree $D_R$ as in~\Cref{thm:bmv-regular}, we first note we may assume $D_R$ is sufficiently large by duplicating the left side several times.
Dinur and Harsha pick a sampler graph $H = (A\cup B,E_H)$ with sides of size $D_R$ and small second singular value; we do the same here, except that we ensure that the degrees are power of $2$. They then take each right vertex $v\in R$, consider its neighbours $\Gamma(v) = \{u_1,\ldots,u_{D_R}\}$, identify it with a copy of $A$ and replace $v$ with a copy of $B$, inserting edges according to $H$ with the constraints inherited from the respective edges in $\Psi$. Dinur and Harsha show that this transformation preserves completeness, and that the increase in the soundness vanishes with the second singular value of $H$. Also, it is easy to see that the right degree of the graph is the degree of $H$, and the left degree is multiplied by the degree of $H$. Summarizing, we get the following result:
\begin{theorem}
\label{thm:bmv-prime}
For every constant $\eps>0$, there is $d>0$, such that for all sufficiently large $t$, there is $C>0$ and a polynomial-time reduction that given $3$-SAT instance $\varphi$ produces an instance $\Psi$ of label cover of size at most $n (\log n)^C$, alphabet size at most $C$, all constraints being $d$-to-$1$, and bi-regular with degrees $O_{t,\eps}(1)$ such that the following holds.
\begin{enumerate}
    \item \textbf{Completeness:} If $\varphi$ is satisfiable, then $\val(\Psi)=1$.

    \item \textbf{Soundness:} If $\varphi$ is unsatisfiable, then $\val(\Psi)\le\eps$.

    \item \textbf{Right degree:} The right degree is equal to $2^t$.
\end{enumerate}
\end{theorem}

\subsection{Agreement Soundness}
We next turn the result of~\Cref{thm:bmv-regular} to a result regarding \emph{agreement soundness}, a notion due to~\cite{Mos15} which is tailored for coverage problems

\begin{definition}
Let $\Psi = ((L\cup R), E,\Sigma_L,\Sigma_R,\{\phi_{e}\}_{e\in E})$ be a label cover instance, and let $A_L\colon L\to\Sigma_L$ be an assignment. We say that the neighbors of $v\in R$ totally disagree under $A_L$ if for every two distinct neighbors $u,u'$ of $v$ we have that $\phi_{(u,v)}(A_L(u))\neq \phi_{(u',v)}(A_L(u'))$.
\end{definition}
The disagreement value of an assignment $A_L$ is the fraction of vertices in $R$ that are in total disagreement under $A_L$. The disagreement value of an instance $\Psi$, denoted by $\mathsf{disagval}(\Psi)$, is the minimum over all $A_L\colon L\to \Sigma_L$ of the disagreement value of $A_L$.  We have the following lemma due to~\cite{Mos15}:
\begin{lemma}\label{lem:mosh}
    Let $D_R$ be a prime power, $t\in\mathbb{N}$, $D=D_R^t$ and let $\eps>0$. There is a polynomial time reduction that maps a label cover instance $\Psi$ which is right-regular with degree $D$, to a label cover instance $\Psi'$ which is right regular with degree $D_R$, with the following properties:
    \begin{enumerate}
    \item \textbf{Completeness:} If $\val(\Psi)=1$, then $\mathsf{val}(\Psi')=1$.

    \item \textbf{Soundness:} If $\val(\Psi)\leq \eps^2 D_R^2$, then $\mathsf{disagval}(\Psi')\geq 1 - 2\eps D_R^2$.

    \item \textbf{Parameters:} The size of $\Psi'$ is larger by factor at most $O_{t,D}(1)$, the alphabets are the same, and if all constraints of $\Psi$ are $d$-to-$1$, then all constraints of $\Psi'$ are $d$-to-$1$.
\end{enumerate}
\end{lemma}
\begin{proof}
The first two items are the same as in~\cite[Lemma 4.4]{Mos15}, and the third item follows immediately by inspecting the construction.    
\end{proof}
We will need the notion of list agreement, which generalizes the notion of agreement soundness, and a lemma about it from~\cite{Mos15}.

\begin{definition}
   Let $\Psi = ((L\cup R), E,\Sigma_L,\Sigma_R,\{\phi_{e}\}_{e\in E})$ be a label cover instance. We say that $\Psi$ has list agreement soundness $(\ell,\eps)$ if for any $A_L\colon L\to \binom{\Sigma_L}{\ell}$, for at least $1-\eps$ fraction of the vertices $v\in R$ it holds that for any distinct neighbours $u,u'$ of $v$ and any $\sigma\in A_L(u)$, $\sigma'\in A_L(u')$, we have that $\phi_{(u,v)}(\sigma)\neq \phi_{(u',v)}(\sigma')$.
\end{definition}

The following result is~\cite[Lemma 4.7]{Mos15}.
\begin{lemma}\label{lem:list_disagree}
    If a label cover instance $\Psi$ has $\mathsf{disagval}(\Psi)\geq 1-\eps$, then it has list agreement soundness $(\ell, \ell^2 \eps)$ for all $\ell\in\mathbb{N}$.
\end{lemma}

Summarizing, we get the following result:
\begin{theorem}\label{thm:list_soundness_form}
For every $\eps>0$, $s\in\mathbb{N}$ there are $d>0$, $C>0$ and a polynomial-time reduction that given $3$-SAT instance $\varphi$ produces an instance $\Psi$ of label cover of size at most $n (\log n)^C$, alphabet size at most $C$, all constraints being $d$-to-$1$, and bi-regular with degrees $O_{\eps,s}(1)$ such that the following holds. 
\begin{enumerate}
    \item \textbf{Completeness:} If $\varphi$ is satisfiable, then $\mathsf{val}(\Psi)=1$.

    \item \textbf{Soundness:} If $\varphi$ is unsatisfiable, then $\Psi$ has list agreement soundness $(\ell,2^s \ell^2 \eps)$ for all $\ell\in\mathbb{N}$.

    \item \textbf{Right degree:} The right degree is equal to $2^s$.
\end{enumerate}   
\end{theorem}
\begin{proof}
    We use the reduction from~\Cref{thm:bmv-prime} with $\eps^2/16$ and $t = sT$ where $T$ is sufficiently large, and then apply on it~\Cref{lem:mosh} with $D_R=2^s$. The completeness is clear, and in the soundness we get that if $\varphi$ is unsatisfiable, then $\mathsf{disagval}(\Psi)\geq 1-4\cdot 2^s\cdot \sqrt{\eps^2/16} = 1-2^s \eps$. The result then follows from~\Cref{lem:list_disagree}.
\end{proof}

\subsection{The Basic Hypergraph Construction}
We next turn the result of~\Cref{thm:list_soundness_form} to a result regarding coverage in hypergraphs using the reduction of~\cite{Feige98}.  More specifically, we establish the following result:
\begin{theorem}\label{thm:apply_feige}
    For every $\eps>0$ there are $C_1,C_2,C_3>0$ and a polynomial time reduction that given a $3$-SAT instance $\varphi$ produces a $C_2$-uniform hypergraph $\HH = (V,\mathcal{E})$ of size at most $n(\log n)^{C_1}$ such that setting $k = |V|/C_2$, we have the following properties:
    \begin{enumerate}
    \item \textbf{Completeness:} If $\varphi$ is satisfiable, then there exists a set of vertices $S\subseteq V$ of size $k$ that touches each hyperedge of $\HH$ in exactly one vertex of $S$.

    \item \textbf{Soundness:} If $\varphi$ is unsatisfiable, then $\mathsf{OPT}_{\HH}(k)\leq 1-\frac{1}{e}+\eps$.
\end{enumerate}
    In both cases, the hypergraph $\HH$ is $C_3$-regular.
\end{theorem}
The rest of this section is devoted to the proof of~\Cref{thm:apply_feige}. Fix a label cover instance $\Psi = ((L\cup R),E,\Sigma_L,\Sigma_R,\{\phi_e\}_{e\in E})$ as in~\Cref{thm:list_soundness_form} for parameters $\eps>0$ and $s\in\mathbb{N}$ to be determined, and say each constraint is $d$-to-$1$. 
We produce the hypergraph $\HH = (V,\mathcal{E})$ as follows:
\begin{itemize}
    \item The vertex set is $V = L\times \Sigma_L$.
    \item For each vertex $v\in R$ and $f\colon \Sigma_R\to [2^s]$, we add a hyperedge $e_{v,f}$ to $\mathcal{E}$ as follows. Let $u_1,\ldots,u_{2^s}$ be the neighbor of $v$ in $\Psi$; then we define
    \[
    e_{v,f} = \left\{(u_i,\sigma)~|~f(\phi_{u_i,v}(\sigma))=i\right\}.
    \]
    \item We take the parameter $k$ to be equal to $|L|$.
\end{itemize}

We now analyze the reduction. We first note that this hypergraph is uniform.
\begin{claim}\label{claim:uniform_feige}
    The graph $\HH$ is a $d|\Sigma_R|$-uniform hypergraph and $k = \frac{|V|}{d|\Sigma_R|}$.
\end{claim}
\begin{proof}
Fix 
$v\in R$ and $f\colon \Sigma_R\rightarrow [2^s]$ and denote $x_i = |f^{-1}(i)|$. Because the constraint $(u_i,v)$ is $d$-to-$1$, for each $\tau\in f^{-1}(i)$ there are $d$ values $\sigma\in \Sigma_L$ such that $\phi_{u_i,v}(\sigma)=\tau$. Thus, the number of $\sigma$ such that $(u_i,\sigma)\in e_{v,f}$ is equal to $d\cdot x_i$, and so $|e_{v,f}|=\sum\limits_{i=1}^{2^s}d\cdot x_i = d |\Sigma_R|$. Lastly, note that:
\[
k = |L| = \frac{|V|}{|\Sigma_L|} = \frac{|V|}{d|\Sigma_R|}.\qedhere
\]
\end{proof}

We next analyze the degree of vertices in $\HH$.
\begin{claim}\label{claim:degree_feige}
    Denoting the left degree of $\Psi$ by $D_L$, the degree of each vertex in $\HH$ is $D_L\cdot 2^{s(|\Sigma_R|-1)}$.
\end{claim}
\begin{proof}
    Fix a vertex $(u,\sigma)$, fix $v$ a neighbour of $u$ in $\Psi$, and say that $u$ is the $i^{\text{th}}$  neighbour of $v$. For $(u,\sigma)$ to appear in the edge $e_{v,f}$ we must have that $f(\phi_{u,v}(\sigma))=i$, and this happens for $(2^s)^{|\Sigma_R|-1}$ of the functions $f\colon \Sigma_R\to [2^s]$. Since $u$ has $D_L$ neighbours in $\Psi$, the result follows.
\end{proof}

The following claim asserts the completeness of the reduction.
\begin{claim}\label{claim:completeness_feige}
    If $\mathsf{val}(\Psi)=1$, then there is a set $S\subseteq V$ of size $k$ that touches each hyperedge of $\HH$ exactly once.
\end{claim}
\begin{proof}
    Let $A_L\colon L\to\Sigma_L$, 
    $A_R\colon R\to\Sigma_R$ be a satisfying assignment for $\Psi$. Define the set 
    \[
    S = \{(u,A_L(u))~|~u\in L\}.
    \]
    Clearly, $|S|=|L|=k$. Next, fix $v\in R$, $f\colon \Sigma_R\to [2^s]$ and take $i = f(A_R(v))$. Note that if $u$ is the $i^{\text{th}}$  neighbour of $v$, then 
$    f(\phi_{u,v}(A_L(u)))
    =f(A_R(v))
    =i$,
    so $(u,A_L(u))\in e_{v,f}$. Thus, the set $S$ touches the hyperedge $e_{v,f}$. For any other neighbour $u'\neq u$ of $v$, we again have that $
    f(\phi_{u',v}(A_L(u')))
    =f(A_R(v))
    =i$,
    but $u'$ is not the $i^{\text{th}}$  neighbour of $v$, so we get that $(u',A_{L}(u'))\not\in e_{v,f}$.
\end{proof}

Finally, we analyze the soundness of the reduction.
\begin{claim}\label{claim:soundness_feige}
    If $\Psi$ has $(\ell, \ell^2 \eps)$ list agreement soundness, then for all $S\subseteq V$ of size $k$ and for all $\ell\geq 1$ we have that
    $\mathsf{cov}_{\HH}(S)\leq 1 - (1-2^{-s})^{2^s}+\ell^2 \eps + \frac{2^s}{\ell}$.
\end{claim}
\begin{proof}
    Let $S\subseteq V$ be a set of size $k$. For each vertex $u\in L$ define the list 
    \[
    L(u) = \left\{\sigma\in \Sigma_L~|~(u,\sigma)\in S\right\}.
    \]
    We get 
    $\sum\limits_{v\in L} |L(v)| = |S| = k = |L|$, 
    so the average size of the list is $1$. We say a vertex $v\in L$ is heavy if $|L(v)|\geq \ell$, and otherwise we say that it is light. Then by an averaging argument there are at most $|L|/\ell$ heavy vertices, and we separately bound the amount of hyperedges covered by heavy elements and by light elements.
    \vspace{-2ex}
    \paragraph{Coverage by heavy elements:} Sampling an edge $(u,v)$ of $\Psi$ uniformly, by bi-regularity we get that the marginal distribution of $u$ is uniform over $L$, and thus it is heavy with probability at most $1/\ell$. Sampling a vertex $v\in R$ and then inspecting all of its edges $(u_1,v),\ldots,(u_{2^s},v)$, we conclude by the union bound that the probability that at least one of the $u_i$ is heavy is at most $\frac{2^s}{\ell}$. This implies that at most $\frac{2^s}{\ell}$ fraction of hyperedges $e_{v,f}$ are covered by 
    an element of the form $(u,\sigma)$ where $u$ is heavy.
    \vspace{-2ex}
    \paragraph{Coverage by light elements:}
    Define
    \[
    L'(u)=
    \begin{cases}
        L(u) & \text{if $u$ is light},\\
        \emptyset & \text{if $u$ is heavy},
    \end{cases}
    \]
    and define the assignment $A_L\colon L\to \binom{\Sigma_L}{\leq \ell}$ by $A_L(u) = L'(u)$. Then by the list agreement soundness, for at least $1-\ell^2 \eps$ fraction of $v\in R$, for any distinct neighbours $u,u'$ of $v$ and all $\sigma\in A_L(u)$, $\sigma'\in A_L(u')$ we have that $\phi_{u,v}(\sigma)\neq \phi_{u',v}(\sigma')$. We call such $v$ typical. Fix a typical $v$, let $u_1,\ldots,u_{2^s}$ be its neighbours, and take a uniformly chosen $f\colon \Sigma_R\to [2^s]$. Then
    \[
    \begin{aligned}
    \Pr_{f\colon\Sigma_R\to[2^s]}[\forall i\in[2^s], \forall \sigma\in L'(u_i), (u_i,\sigma)\not\in e_{v,f}]
    =
    \prod\limits_{i=1}^{2^s}
    \left(1-2^{-s}\right)^{|\phi_{u_i,v}(L'(u_i))|}
    \geq \left(1-2^{-s}\right)^{\sum\limits_{i=1}^{2^s}|L'(u_i)|},
    \end{aligned}
    \]
    where in the first equality, we used the fact that for $i<i'$, $\{\phi_{u_i,v}(\sigma)~|~\sigma\in L'(u_i)\}$ and $\{\phi_{u_{i'},v}(\sigma)~|~\sigma\in L'(u_{i'})\}$ are disjoint, so the values $f$ receives on them are independent. Denoting $X_v = \sum\limits_{i=1}^{2^s}|L'(u_i)|$, the lower bound we get can be expressed as $(1-2^{-s})^{X_v}$. Overall, the fraction of hyperedges covered by light elements is therefore at most
    \[
    \mathop{\mathbb{E}}_{v\in R}[1_{v\text{ is not typical}}]
    +\mathop{\mathbb{E}}_{v\in R}[1_{v\text{ is typical}}(1-(1-2^{-s})^{X_v})]
    \leq 
    \ell^2 \eps
    +1-\mathop{\mathbb{E}}_{v\in R}[(1-2^{-s})^{X_v}],
    \]
    which using Jensen's inequality is at most 
$    \ell^2\eps+1-(1-2^{-s})^{\mathop{\mathbb{E}}_{v\in R}[X_v]}$.
    Noting that 
    \[
    \mathop{\mathbb{E}}_{v\in R}[X_v]
    =\mathop{\mathbb{E}}_{v\in R}\left[2^s\mathop{\mathbb{E}}_{u\text{ neighbour of }v}[|L'(u)|]\right]
    \leq 2^s,
    \]
    we get that the fraction of hyperedges covered by light elements is at most $1 - (1-2^{-s})^{2^s}+\ell^2 \eps$.
\end{proof}

We now complete the proof of~\Cref{thm:apply_feige}:
\begin{proof}[Proof of~\Cref{thm:apply_feige}]
    Fix $\eps>0$ and take large $s$ so that $(1-2^{-s})^{2^s}\geq \frac{1}{e}-\frac{\eps}{3}$. Take $\eta = \frac{\eps^3}{27\cdot 2^{3s}}$, take the label cover instance from~\Cref{thm:list_soundness_form} with parameters $\eps=\eta$ and $s$, and then $(\HH,k)$ as described above.
    The uniformity and degree follow from~\Cref{claim:degree_feige} and ~\Cref{claim:uniform_feige}. Then the completeness case follows from~\Cref{claim:completeness_feige}, and for the soundness we have from~\Cref{claim:soundness_feige} that
    $\mathsf{OPT}_{\HH}(k)\leq 1 - (1-2^{-s})^{2^s}+\ell^2 \eta + \frac{2^s}{\ell}$. Plugging in $\ell = \frac{3\cdot 2^{s}}{\eps}$ and elementary manipulations gives that $\mathsf{OPT}_{\HH}(k)\leq 1-\frac{1}{e}+\eps$.
\end{proof}

\subsection{The Blown-up Hypergraph Construction}
We are now ready to complete the proof of~\Cref{thm:bmv-hypergraph-source}.
\begin{proof}[Proof of~\Cref{thm:bmv-hypergraph-source}]
    Fix $\eps>0$ and take the hypergraph $\HH = (V,\mathcal{E})$ from~\Cref{thm:apply_feige} with the constants $C_1,C_2,C_3$ from there. We define the hypergraph $\HH'=(V\times [\log^3 n], \mathcal{E}')$ where for each hyperedge $e = \{v_1,\ldots,v_{C_2}\}\in \mathcal{E}$ we include in $\HH'$ all hyperedges $\{(v_1,i_1),\ldots,(v_{C_2},i_{C_2})\}$, and define $k' = k\log^3 n$. First, as the size of $\HH$ is at most $n\log^{C_1} n$, the size of $\HH'$ is at most $N = n\log^{C_1+3C_2} n$.
    Second, since $\HH$ is $C_3$-regular, the hypergraph $\HH'$ is regular with degree $\Delta = C_3 \log^{3(C_2-1)} n$. Third, the codegree of $\HH'$ is at most $C_3\log^{3(C_2-2)} n = \Delta/\log^3 n\leq \Delta/ \log^2 N$ for all sufficiently large $n$.

    As per the completeness, if $\varphi$ is satisfiable, then we may find $S\subseteq V$ of size $k$ touching each hyperedge of $\HH$ exactly once. Defining 
    $S' = \{(u,j)~|~u\in S, j\in [\log ^3 n]\}$, we get that $S'$ has size $k'$ and touches each hyperedge of $\HH'$ exactly once.

    Lastly, for the soundness, suppose $S'\subseteq V\times [\log^3 n]$ has size $k'$. For each $v\in V$ define $\alpha_v = \Pr_{j\in [\log^3 n]}[(v,j)\in S']$, so that we get that 
    $\sum\limits_{v\in V}\alpha_v = k$.
    Note that for any hyperedge $e = \{v_1,\ldots,v_{C_2}\}\in \mathcal{E}$, the fraction of hyperedges in $\HH'$ that $e$ produces covered by $S'$ is exactly \mbox{$
    1-\prod\limits_{i=1}^{C_2}(1-\alpha_{v_i})$},
    so the total fraction of hyperedges covered by $S'$ is equal to 
    $1-\mathop{\mathbb{E}}_{\{v_1,\ldots,v_{C_2}\}\in \mathcal{E}}[\prod\limits_{i=1}^{C_2}(1-\alpha_{v_i})]$, and we think of it as a function $F(\alpha_{x_1},\ldots,\alpha_{x_{n'}})$ over the domain 
    \[
    \left\{(\alpha_{x_1},\ldots,\alpha_{x_{n'}})~|~\alpha_{x_i}\in [0,1]~\forall i\in[n'], \sum\limits_{i\in[n']}\alpha_{x_i} = k\right\},
    \]
    where $n' = |V|$. Defining the function $f\colon \{S\subseteq V\}\to [0,1]$ by $f(S) = \mathsf{cov}_{\HH}(S)$, we see that $f$ is monotone sub-modular and $F$ is the multilinear extension of $f$. Thus, by~\cite[Lemma 3.5]{CalinescuChekuriPalVondrak2011} there is $S$ of size $k$ such that
    \[
    F(\alpha)\leq f(S) = \mathsf{cov}_{\HH}(S)
    \leq 1-\frac{1}{e}+\eps,
    \]
    as required. 
\end{proof}

\section{Some Hypergraph Results}\label{sec:explicit-colorings}
In this section, we first provide an explicit family of hypergraph colorings and second provide a structural result on hypergraphs, both of which will be used in our reduction in the next section.
\subsection{An Explicit Family of Colorings}

In this section, we show how to construct a family of vertex coloring of hypergraphs in quasipolynomial time with certain nice properties that are useful for the reduction to clustering problems in the subsequent sections.
We will use the following standard concentration bound for sums of bounded-independence
random variables.

\begin{proposition}
\label{lem:bounded-independence-concentration}
Let $r\geq2$ be even, and let $X_1,\ldots,X_N\in[0,B]$ be
$r$-wise independent random variables. Let $X=\sum_{i\in[N]}X_i$ and
$\sigma^2=\sum_{i\in[N]}\Var(X_i)$. There is an absolute constant $C$ such
that
\begin{align*}
\E\bigl[|X-\E[ X]|^r\bigr]
\leq C^r\bigl(\sqrt{r\sigma^2}+rB\bigr)^r \quad\text{and}\quad
\Pr\bigl[|X-\E[ X]|>4C(\sqrt{r\sigma^2}+rB)\bigr]
\leq4^{-r}.
\end{align*}
\end{proposition}

\begin{proof}
Expanding $(X-\E [X])^r$ gives terms involving at most $r$ distinct
variables. By $r$-wise independence, its expectation is therefore the
same as for fully independent variables with the same marginals. The
first bound follows from the moment form of Bernstein's inequality,
and the second from Markov's inequality.
\end{proof}

\begin{theorem}
\label{lem:enumerable-good-colorings}
Fix an integer $t\geq2$, and let $n,k,d,\chi$ be sufficiently large integers such that
$2td\sqrt n\leq k=o(n/\log n)$. Then, in time
$2^{O((\log n)^2)}$ one can construct a list $\mathcal C_{n,k}$ of
$k$-colorings of $[n]$ with the following property.
If $\mathcal{H}$ is a $t$-uniform hypergraph over $[n]$ with degree $d$ and codegree $\chi$, then there exists $\rho\in \mathcal{C}_{n,k}$ with the following properties:
\begin{enumerate}
\item \textbf{Nearly balanced colors:} For every $c\in[k]$, we have 
$\left||\rho^{-1}(c)|-\frac nk\right|
=o(n/k)$.
\item \textbf{Nearly all hyperedges rainbow:} At most
$100\binom{t}{2}\frac{nd}{tk}$ hyperedges have two vertices of the same color.
\item \textbf{Neighbors of one color:} For every $v\in[n]$ and
$c\in[k]$, at most three vertices $u\neq v$ have color $c$ and
appear together with $v$ in a hyperedge.
\item \textbf{Hyperedges using two colors:} For every pair of distinct
colors $a,b\in[k]$, at most $120\chi\lceil\log_2 k\rceil$ rainbow
hyperedges use both colors.
\item \textbf{Balanced with Weights:} For any fixed set of numbers $\{q_v\in[0,d]\mid v\in [n]\}$ satisfying
$\sum_{v\in[n]}q_v=nd/t$, the coloring can be chosen to satisfy
the above properties as well as
\begin{align*}
\left|\sum_{v:\rho(v)=c}q_v-\frac{nd}{tk}\right|
=o\left(\frac{nd}{tk}\right)
\end{align*}
for every $c\in [k]$.
\end{enumerate}
\end{theorem}
\begin{proof}
Setting $r=20\lceil\log_2 k\rceil$, we take $\mathcal C_{n,k}$ an
$r$-wise independent family of functions $\rho:[n]\to[k]$ with uniform
marginals; the explicit construction of~\cite{AlonBabaiItai1986} achieves near
uniform marginals (and would suffice for us), but it is more convenient
to use the construction of~\cite{HarveySahami2024} that achieves perfectly uniform
marginals. Both have size $n^{O(r)}=2^{O((\log n)^2)}$.

Fixing a hypergraph $\mathcal H$, we prove that a randomly chosen
$\rho\in\mathcal C_{n,k}$ satisfies all five properties with probability
close to $1$. Using Proposition~\ref{lem:bounded-independence-concentration} with the constant $C$ from there and the union
bound, the probability that
\[
\left||\rho^{-1}(c)|-\frac nk\right|
>
4C\left(\sqrt{\frac{rn}{k}}+r\right)
= o(n/k)
\]
for some
$c\in[k]$ is at most $k\cdot 4^{-r}\leq \frac1{100}$.

For the second item, for each hyperedge $e$ let $Y_e$ be the indicator
random variable that it is not rainbow under $\rho$. Then by
pairwise independence $\mathbb E[Y_e]\leq \binom{t}{2}\frac{1}{k}$,
so, using $|E|=nd/t$,
$\mathbb E\left[\sum_{e\in E}Y_e\right]
\leq
\binom{t}{2}\frac{nd}{tk}$.
Markov's inequality now gives that the second item fails with probability
at most $1/100$.

For the third item, fix $v\in[n]$ and $c\in[k]$. There are at most $td$
vertices $u$ that appear with $v$ in some hyperedge of $\mathcal H$.
Fixing any $4$ such vertices, the probability they are all colored $c$
is $1/k^4$ by $4$-wise independence, so the probability there are at
least $4$ such vertices is at most $\binom{td}{4}\frac1{k^4}$.
Using the union bound, the probability the third item fails is at most
\[
nk\binom{td}{4}\frac1{k^4}
\leq
\frac{(td)^4n}{k^3}
\leq
\frac{k}{16n}
\leq
\frac1{100},
\]
where in the second transition we used $k\geq 2td\sqrt n$ and in the
third one we used $k=o(n/\log n)$.

For the fourth item, assume the third item holds and fix distinct
$a,b\in[k]$. Suppose we have hyperedges $\mathcal L$ of size
$R\geq120\chi\lceil\log_2k\rceil$ in $\mathcal H$ that are rainbow
and use colors $a,b$. We may choose a subcollection
$\{e_1,\ldots,e_{R'}\}\subseteq\mathcal L$ such that no vertex appearing
in more than one of the $e_i$ is colored either $a$ or $b$, as follows:
\begin{enumerate}
\item Start with $\mathcal L'=\emptyset$.
\item As long as $\mathcal L\neq\emptyset$, pick a hyperedge
$e'\in\mathcal L$ arbitrarily from it, add it to $\mathcal L'$, and let
$u,v\in e'$ be the $a$-colored and $b$-colored vertices in $e'$. By the
third item, there are at most $6$ vertices $x$ that are colored $a$ or
$b$ and appear in a hyperedge with $u$. For each such $x$ we remove from
$\mathcal L$ all hyperedges containing both $u$ and $x$. We do the same
for $v$, so overall we remove from $\mathcal L$ at most $12\chi$ edges.
\end{enumerate}
It is clear that when the process terminates, we get a collection as
required and that $R'\geq \frac{R}{12\chi}$.
In particular, we can find $10\lceil\log_2k\rceil$ hyperedges that are
colored as specified above.

On the other hand, the probability that for a randomly chosen $\rho$
there will be such a configuration of hyperedges, for fixed $a,b$, is
at most
\[
\left(\frac{nd}{t}\right)^{10\lceil\log_2k\rceil}
\left(\frac{t^2}{k^2}\right)^{10\lceil\log_2k\rceil}
\leq
\frac{1}{2^{10\lceil\log_2k\rceil}}
\leq
\frac1{k^{10}},
\]
so the probability there are some colors $a,b\in[k]$ obtaining this
configuration is at most
\[
k^2\frac1{k^{10}}\leq\frac1{100}.
\]
We conclude that the probability the fourth item fails is at most the
probability that the third item fails plus $1/100$, hence is at most
$2/100$.

For the fifth item, fix numbers $\{q_v\}_{v\in[n]}$ as in the statement.
For a fixed color $c\in[k]$, let $X_v$ indicate that $\rho(v)=c$. The
variables $q_vX_v$ are $r$-wise independent, take values in $[0,d]$,
and
\[
\mathbb E\left[\sum_{v\in[n]}q_vX_v\right]
=
\frac1k\sum_{v\in[n]}q_v
=
\frac{nd}{tk},
\qquad
\sum_{v\in[n]}\Var(q_vX_v)
\leq
\frac1k\sum_{v\in[n]}q_v^2
\leq
\frac{nd^2}{tk}.
\]
Thus, by Proposition~\ref{lem:bounded-independence-concentration} , except with probability
$4^{-r}$,
\[
\left|
\sum_{v:\rho(v)=c}q_v-\frac{nd}{tk}
\right|
\leq
4Cd\left(\sqrt{\frac{rn}{tk}}+r\right)
=
o\left(\frac{nd}{tk}\right).
\]
Taking a union bound over the colors, the probability that the fifth
item fails is at most $k4^{-r}\leq\frac1{100}$.

Overall, by the union bound the probability that the chosen $\rho$ fails to satisfy one of the items is at most $\frac{6}{100}<1$, hence there is $\rho\in\mathcal{C}_{n,k}$ as required.
\end{proof}

\subsection{A Bound on the Sum of Squares of Degrees}
We will need the following result about hypergraphs.
\begin{lemma}
\label{lem:hypergraph-second-moment}
Let $\HH=(V,E)$ be a $t$-uniform hypergraph of maximum
degree at most $d$ and maximum codegree at most $\chi$. Let
$\mathcal F\subseteq E$ be a collection of $m$ hyperedges and for every $v\in V$ denote
$f_v:=|\{e\in\mathcal F:v\in e\}|$.
Then
\[
\sum_{v\in V}f_v^2
\leq
md+O_t\left(\sqrt{\chi}\,m^{3/2}\right).
\]
\end{lemma}
\begin{proof}
For every $e\in\mathcal F$, order its vertices
$v_1(e),\ldots,v_t(e)$ so that $f_{v_1(e)}\geq\cdots\geq f_{v_t(e)}$. Thus, by double counting $\sum_{v\in V}f_v^2
=
\sum_{e\in\mathcal F}\sum_{i=1}^{t}f_{v_i(e)}$.
Since every vertex has degree at most $d$, $\sum_{e\in\mathcal F}f_{v_1(e)}\leq md$, and we now show that the contribution of $i>1$ is dominated by the error term.

Fix $i>1$. Defining $
S_j:=\{v\in V:f_v\geq j\}$, we get by an averaging argument that $|S_j|\leq\frac{tm}{j}$.
If $f_{v_i(e)}\geq j$, then $e$ contains at least two vertices of
$S_j$. Since every pair of vertices belongs to at most $\chi$
hyperedges, the number of such hyperedges is at most
\[
\min\left\{
m,\,
\chi\binom{|S_j|}{2}
\right\}
\leq
\min\left\{
m,\,
\frac{\chi t^2m^2}{2j^2}
\right\}.
\]
Therefore
\[
\sum_{e\in\mathcal F}f_{v_i(e)}
\leq
\sum_{j\geq1}
\min\left\{
m,\,
\frac{\chi t^2m^2}{2j^2}
\right\}
=
\sum_{j < \sqrt{\chi m}}
m
+
\sum_{j\geq \sqrt{\chi m}}
\frac{\chi t^2m^2}{2j^2}
=
O_t\left(\sqrt{\chi}\,m^{3/2}\right).
\]
Summing over $i=2,\ldots,t$ gives
$\sum\limits_{e\in\mathcal{F}}\sum\limits_{i>1}f_{v_i(e)}\leq O_{t}(\sqrt{\chi}m^{3/2})$, completing the proof.
\end{proof}

\section{Lower Bound for $k$-means under Almost ETH}
\label{sec:kmeans-lower-bound}

The main goal of this section is to prove~\Cref{thm:intro-bmv}, restated below. Fix $\eps = \frac{1}{100}$ in~\Cref{thm:bmv-hypergraph-source}, let $t$ be from that theorem and set $L = 3(t-1)$ throughout this section.

\begin{theorem}
\label{thm:bmv-kmeans-formal}
Assuming Almost ETH, the following holds for every constant $\beta\in(0,1)$. Let $\tilde{\eps}\colon \mathbb{N}\to (0,1)$, $\tilde{k}\colon \mathbb{N}\to\mathbb{N}$ be polynomial time computable functions satisfying:
\begin{enumerate}
    \item $\frac{\tilde{k}(n)}{\tilde{\eps}(n)} = o\left(\frac{n}{\log^L n}\right)$ and the sequence $\frac{\tilde{k}(n)}{\tilde{\eps}(n)}$ has bounded consecutive ratios.
    \item $\omega(\log n)\leq \frac{1}{\tilde{\eps}(n)}\leq o\left(\frac{\tilde{k}(n)}{\log^{2L} n }\right)$.
\end{enumerate}
Then for sufficiently large $n$, setting $\eps = \tilde{\eps}(n)$ and $k = \tilde{k}(n)$, Euclidean
$k$-means on $n$ points in $\mathbb R^d$ has no
$(1+\varepsilon)$-approximation algorithm running in time    $2^{(k/\varepsilon)^{1-\beta}}
    \operatorname{poly}(n,d)$.
\end{theorem}

The proof of~\Cref{thm:bmv-kmeans-formal} is based on a reduction from hypergraph coverage as follows.
\begin{theorem}
\label{thm:hypergraph-to-kmeans}
Let $\HH=(V, E)$ be a $t$-uniform $d$-regular hypergraph on $N$ vertices where $d = d(N)$ grows with $N$, let  $k$ be an integer satisfying
$2td\sqrt N\leq k=o(N/\log N)$, and suppose that the codegree of $\HH$ is at most
$o(d/\log k)$. 
Then there is an algorithm running in time
$2^{O(\log^2 N)}$ that transforms $\HH$ into a collection $\mathcal{P}$ of
at most $2^{O(\log^2 N)}$ subsets of $\{x\in\{0,1\}^{N+k}~|~|x|=2t\}$ of size at most $M=Nd/t$ each, such that the following holds.
\begin{enumerate}
    \item \textbf{Completeness:} If there is $S\subseteq V$ of size $N/t$ that touches each hyperedge of $\HH$ once, then there is $P\in\mathcal{P}$ such that
    \[
        \OPT^{2}_k(P)\leq(2t-1)M-(1-o(1))kd.
    \]

    \item \textbf{Soundness:} If every $S\subseteq V$ of size $N/t$ fails to cover at least $M/4$ of the hyperedges of $\HH$, then for all $P\in\mathcal{P}$ we have
    \[
        \OPT^{2}_k(P)\geq(2t-1)M-\left(1-\frac{1}{8t}\right)kd.
    \]
\end{enumerate}
\end{theorem}
We first show that~\Cref{thm:hypergraph-to-kmeans} implies
\Cref{thm:bmv-kmeans-formal} via a padding reduction. In~\Cref{thm:hypergraph-to-kmeans}, the gap between the YES case and the NO case is an additive factor of $\Theta\left(\frac{kd}{t}\right)$ while the cost itself is $\Theta(tM)$, so if they were comparable the additive gap would already be a multiplicative gap as needed in~\Cref{thm:bmv-kmeans-formal}. To arrange that, the idea is to pick the first $q\geq N$ where $\frac{k(q)}{\eps(q)}\geq B\cdot tN$, and then duplicate points in the construction given in~\Cref{thm:hypergraph-to-kmeans} to get $q$ points in each set.

\begin{proof}
Let $\HH=(V,E)$ be an instance from~\Cref{thm:bmv-hypergraph-source}
with $\eps=1/100$, let $N$, $t$ and $d$ denote the number of vertices, uniformity and degree of $\HH$ respectively, and let $M=Nd/t$. Let $B$ be a sufficiently large constant and fix $q$ to be the first integer
such that $\frac{\widetilde k(q)}{\widetilde\eps(q)}
    \geq BtN$. As $\frac{\widetilde k(N)}{\widetilde\eps(N)}=o(N)$ we get that $q\geq N$, so by the bounded consecutive ratio assumption we have $\frac{\widetilde k(q)}{\widetilde\eps(q)}
    =\Theta(N)$.
    
Setting $k = \widetilde{k}(q)-1$ and $\eps = \widetilde{\eps}(q)$, we get that $\HH$ satisfies the assumptions of~\Cref{thm:hypergraph-to-kmeans}, and applying it we get a collection $\mathcal{P}$ of $2^{O(\log^2 N)}$ subsets as therein. For each $P\in\mathcal{P}$ we have 
\[
|P|\leq M = \frac{Nd}{t} \leq O(N\log^{L} N)
\leq O\left(\frac{\tilde{k}(q)\log^{L}N}{\tilde{\eps}(q)}\right) = o\left(\frac{q\log^{L}N}{\log^{L}q}\right)=o(q).
\]
Thus, letting
$W:=\lfloor(q-1)/|P|\rfloor$, we form the set $P'$ by taking $W$ copies of every point of $P$ where we append one zero on a new coordinate $\mathbf e_{\mathrm{new}}$. We also add to it $q-W|P|$ copies of
$x:=4tq\mathbf e_{\mathrm{new}}$, so overall $|P'|=q$.

We note that for each $P$ it holds that $\OPT_{\widetilde k(q)}^{2}(P')
    =
    W\OPT_k^{2}(P)$. Indeed, any cluster containing both $x$ and an original point
$p$ has cost at least $\|x-p\|_2^2/2>2tq$, whereas placing all copies
of $x$ in one cluster and all original points at the origin has cost at
most $2tq$. Thus, in an optimal solution one center must be devoted to the copies of $x$ (hence they contribute $0$), while
the remaining $k$ centers cluster the duplicated copy of $P$.  

Denoting $S = (2t-1)M-\left(1-\frac{1}{8t}\right)kd$, in the NO case we get that 
$\OPT^{2}_{\tilde{k}(q)}(P')\geq W S$ for all $P'$, and 
in the YES case we get that there is $P'$ such that 
\[
\OPT^{2}_{\tilde{k}(q)}(P') \leq 
WS-\left(\frac{1}{8}-o(1)\right)W\frac{kd}{t}
\leq
WS-\frac{1}{32}W\frac{BtNd}{t}\eps
= WS - W\frac{B}{32}tM\eps
\leq (1-\eps)WS
\]
for sufficiently large $B$. Thus, if we had an algorithm solving the problem in~\Cref{thm:bmv-kmeans-formal} running in time $2^{(k/\eps)^{1-\beta}}$ for some constant $\beta\in(0,1)$, we could run it on each $P'$ and accept only if one of the runs returned value at most $WS$. This would have been an algorithm running in time $2^{O(N^{1-\beta}+\log^2 N)}\operatorname{poly}(q)$ that distinguishes between the YES and NO case of~\Cref{thm:bmv-hypergraph-source}, which contradicts Almost ETH.
\end{proof}

In the rest of this section we prove~\Cref{thm:hypergraph-to-kmeans}. In~\Cref{sec:hypergraph-kmeans-construction} we construct the point sets and give the completeness, and in~\Cref{sec:hypergraph-kmeans-soundness} we analyze the soundness.

\subsection{The $k$-means Instance and Its Completeness}
\label{sec:hypergraph-kmeans-construction}

Fix a $t$-uniform $d$-regular hypergraph
$\HH=(V,E)$ on $N$ vertices as in
\Cref{thm:hypergraph-to-kmeans}, and fix the family of colorings $\mathcal{C}_{N,k}$  from~\Cref{lem:enumerable-good-colorings}. We drop from $\mathcal{C}_{N,k}$ any coloring failing one of the first four items (this can be done as each one of them can be checked in in polynomial time), and by~\Cref{lem:enumerable-good-colorings} we still have $\mathcal{C}_{N,k}\neq\emptyset$.

For each coloring $\rho\in\mathcal{C}_{N,k}$, we construct a hypergraph $\HH_{\rho} = (V,E_{\rho})$ by deleting from $\HH$ all hyperedges $e\in E$ in which two vertices are colored by the same color under $\rho$;  we refer to such hyperedges as bad. Then by the second item in~\Cref{lem:enumerable-good-colorings} we have that 
\begin{equation}\label{eq:most_edges_are_good}
|E_{\rho}|
\geq M - 100\binom{t}{2}\frac{M}{k}.
\end{equation}
For each $\rho\in\mathcal{C}_{N,k}$, we construct $P_{\rho}\subseteq \{0,1\}^{N+k}$, by including in it the point
\[
p_e:=\sum_{v\in e}\mathbf e_v+
\sum_{c\in\rho(e)}\mathbf e_{N+c}
\]
for each $e\in E_{\rho}$. Note that each $p_e$ has support of size $2t$. 

\vspace{2ex}
This completes the description of the reduction, and to analyze it we need the following well known fact giving us a formula for the cost of a single cluster.
\begin{fact}
\label{lem:folklore}
For a nonempty finite multiset $A\subseteq\mathbb R^d$, the center
minimizing its $1$-means cost is
$c=|A|^{-1}\sum_{x\in A}x$, and
\begin{align*}
 \min_{c\in\mathbb R^d}\sum_{x\in A}\|x-c\|_2^2
 =\frac{1}{2|A|}\sum_{x,y\in A}\|x-y\|_2^2.
\end{align*}
\end{fact}
The following claim gives the completeness of the reduction.
\begin{claim}
    Suppose that there is $S\subseteq V$ of size $N/t$ that touches each hyperedge of $\HH$ exactly once. Then for some $\rho\in\mathcal{C}_{N,k}$ we have that $\OPT_k^{2}(P_{\rho})\leq (2t-1)M-(1-o(1))kd$.
\end{claim}
\begin{proof}
    For every hyperedge $e$ we denote by $v(e)$  the unique vertex in $e\cap S$, and say that $e$ is assigned to $v(e)$. Note that since $\HH$ is $d$-regular, each vertex in $S$
gets assigned exactly $d$ hyperedges (before the bad hyperedges are deleted).
We would like to cluster points $e$ according to the color of $v(e)$, but we first need to ensure the clusters are roughly balanced in size.
Towards this end define $q_v=d$ for $v\in S$ and $q_v=0$
otherwise, so that $q_v$ is precisely the number of hyperedges initially
assigned to $v$ and $\sum_{v\in V} q_v=M$. Choose $\rho\in\mathcal{C}_{N,k}$ that
also satisfies the last conclusion of~\Cref{lem:enumerable-good-colorings}
for these weights, and define the clusters
$\mathcal F_c:=\{e\in E_\rho:\rho(v(e))=c\}$ for $c\in[k]$.
Letting $q'_v$ be the number of not-bad hyperedges assigned to $v$, we have that $q'_v\leq q_v$ for every $v$ so that for all $c\in[k]$
\begin{equation}\label{eq:completeness_of_triv2}
|\mathcal{F}_c|=\sum_{v\in U_c}q'_v
\leq\sum_{v\in U_c}q_v
\leq(1+o(1))\frac{M}{k}.
\end{equation}
Applying~\Cref{lem:folklore}, we get that
\begin{equation}\label{eq:completeness_of_triv}
\cost(\mathcal{F}_c)
=\sum\limits_{e\in\mathcal{F}_c}
\left\|p_e-\frac{1}{|\mathcal{F}_c|}\sum\limits_{e'\in\mathcal{F}_c} p_{e'}\right\|_2^2
= 2t |\mathcal{F}_c| - |\mathcal{F}_c|^{-1}\|\sum\limits_{e\in\mathcal{F}_c}p_e\|_2^2,
\end{equation}
and we now calculate $\|\sum\limits_{e\in\mathcal{F}_c}p_e\|_2^2$. 
\vspace{-1ex}
\paragraph{The contribution of vertex coordinates:} denoting $d_{\mathcal{F}_c,v} = |\{e\in \mathcal{F}_c~|~v\in e\}|$, the contribution of vertex coordinates to $\|\sum\limits_{e\in\mathcal{F}_c}p_e\|_2^2$ is $\sum\limits_{v\in V}d_{\mathcal{F}_c,v}^2
\geq
\sum\limits_{v:\rho(v)=c}(q_v')^2$, where we used the fact that $d_{\mathcal{F}_c,v}\geq q_v'$ for $v$ such that $\rho(v)=c$.

\vspace{-1ex}
\paragraph{The contribution of color coordinates:} denoting $r_{\mathcal{F}_c,a} = |\{e = \{v_1,\ldots,v_t\}\in \mathcal{F}_c~|~a\in \{\rho(v_1),\ldots,\rho(v_t)\}\}|$, the contribution of color coordinates to $\|\sum\limits_{e\in\mathcal{F}_c}p_e\|_2^2$ is 
$\sum\limits_{a\in [k]}r_{\mathcal{F}_c,a}^2\geq r_{\mathcal{F}_c,c}^2 = |\mathcal{F}_c|^2$. 
\vspace{-1ex}
\paragraph{Combining the bounds:} plugging both bounds to~\eqref{eq:completeness_of_triv} gives
\[
\cost(\mathcal{F}_c)\leq (2t-1)|\mathcal{F}_c|-\frac{1}{|\mathcal{F}_c|}\sum\limits_{v:\rho(v)=c}(q_v')^2
\leq (2t-1)|\mathcal{F}_c|-(1-o(1))\frac{k}{M}\sum\limits_{v:\rho(v)=c}(q_v')^2,
\]
where we used~\eqref{eq:completeness_of_triv2}. Summing over $c$ gives 
\[
\OPT^{2}_k(P_{\rho})
\leq (2t-1)\sum\limits_{c}|\mathcal{F}_c| - (1-o(1))\frac{k}{M}\sum\limits_{v}(q_v')^2.
\]
Note that $\sum\limits_{v\in V}q_v^2 = d^2|S|=\frac{N}{t}d^2$. As $q_v\leq d$ for all $v$, removing a single edge from $\HH$ drops this sum by at most $2d-1$, and iterating gives
$\sum\limits_{v\in V}(q_v')^2\geq \frac{N}{t}d^2 - 100\binom{t}{2}\frac{M}{k}\cdot (2d-1)$. Plugging this above and using~\eqref{eq:completeness_of_triv2} gives
\[
\OPT^{2}_k(P_{\rho})
\leq (2t-1)M
-(1-o(1))kd+200(1+o(1))\binom{t}{2}d
\leq (2t-1)M
-(1-o(1))kd.
\qedhere
\]
\end{proof}

\subsection{Soundness Analysis}
\label{sec:hypergraph-kmeans-soundness}
We now analyze the soundness of the construction from~\Cref{sec:hypergraph-kmeans-construction}, and assume henceforth that $\HH$ is as in the NO case. Fixing $\rho\in\mathcal{C}_{N,k}$, we show that if $\OPT^{2}_{k}(P_{\rho})\leq (2t-1)M-(1-1/(8t))kd$, then one could find a set $S\subseteq V$ of size $N/t$ covering more than $\frac{3}{4}M$ of the hyperedges of $\HH$. The goal is to show that any low-cost clustering of $P_{\rho}$ must basically look like the clustering we constructed in the completeness case, and we establish it via the following steps: 
\begin{enumerate}
    \item \textbf{Dominant color in each cluster:} we first establish that in a low-cost clustering of $P_{\rho}$, each cluster must have a dominant color that appears in almost all of the edges $e$ such that $p_{e}\in P_{\rho}$. 
    \item \textbf{Assigning edges to vertices:} the dominant color in each clusters identifies a distinguished vertex $v(e)$ for (almost all) hyperedges $e\in E$, essentially reversing the association from the completeness analysis. We argue that the collection of all distinguished elements is not too spread out, in the sense that many hyperedges $e$ in a cluster pick that same vertex $v$.
    
    \item \textbf{Choosing the set $S$:} finally, we pick our set $S$ to be the collection of $N/t$ vertices that were assigned to the most number of hyperedges, and we show that the previous item implies that $S$ touches many of the hyperedges of $\HH$.
\end{enumerate}

We now move on to the formal proof. Fix a clustering $\mathcal F_1,\ldots,\mathcal F_k$ of $P_\rho$ of cost smaller than $(2t-1)M-\left(1-\frac1{8t}\right)kd$, and assume without loss of generality they are all nonempty. For every $i\in[k]$ define
\[
m_i:=|\mathcal F_i|,
\qquad
d_{i,v}:=|\{e\in\mathcal F_i~|~v\in e\}|,
\qquad
D_i:=\frac1{m_i}\sum_{v\in V}d_{i,v}^2.
\]
Let $\gamma_i$ be the fraction of unordered pairs $\{e,f\}\subseteq\mathcal F_i$ such that $\rho(e)\cap\rho(f)=\emptyset$, and let
\[
\kappa_i:=
\frac1{\binom{m_i}{2}}
\sum_{\{e,f\}\subseteq\mathcal F_i}
\max(|\rho(e)\cap\rho(f)|-1,0).
\]
The following claim expresses the cost of each cluster in terms of the parameters above.
\begin{claim}\label{eq:soundness-cluster-cost}
We have $\cost(\mathcal F_i)
=
(2t-1)m_i-(t-1)-z_i$ where  $z_i:=D_i-(m_i-1)\gamma_i+(m_i-1)\kappa_i$.
\end{claim}
\begin{proof}
Writing $s_i = \sum\limits_{e\in \mathcal{F}_i}p_e$,~\Cref{lem:folklore} gives us that
\[
\cost(\mathcal{F}_i) = 
\sum\limits_{e\in\mathcal{F}_i}\|p_e\|_2^2
-\frac{1}{m_i}\|s_i\|_2^2
=2tm_i-\frac{1}{m_i}\|s_i\|_2^2,
\]
and the goal is to calculate $\|s_i\|_2^2$. Using the notation $d_{i,v}$ above and letting $r_{i,c}$ be the number of hyperedges of $\mathcal F_i$ containing color $c$, we have that 
 $\|s_i\|_2^2 = \sum\limits_{c\in [k]}r_{i,c}^2+\sum\limits_{v\in V}d_{i,v}^2$, and we now calculate each one of these sums. For the first one we have
\[
\sum_{c\in[k]}r_{i,c}^2
=
\sum_{c\in[k]}r_{i,c}
+
2\sum_{c\in[k]}\binom{r_{i,c}}2
=
tm_i+
2\sum_{\{e,f\}\subseteq\mathcal F_i}
|\rho(e)\cap\rho(f)|
=tm_i+
2\binom{m_i}{2}(1-\gamma_i+\kappa_i).
\]
and therefore $\sum_{c\in[k]}r_{i,c}^2
=
tm_i+m_i(m_i-1)(1-\gamma_i+\kappa_i)$. For the second one we have $\sum_{v\in V}d_{i,v}^2=m_iD_i$, and combining and plugging above gives
\[
\cost(\mathcal{F}_i)
=2tm_i - t - (m_i-1)(1-\gamma_i+\kappa_i) - D_i,
\]
and rearranging gives the claim.
\end{proof}
Summing~\Cref{eq:soundness-cluster-cost}, we get the following result.
\begin{claim}\label{claim:sum_of_z}
    $\sum_{i=1}^k z_i\geq \left(1-\frac1{8t}-o(1)\right)kd$.
\end{claim}
\begin{proof}
    Summing Claim~\ref{eq:soundness-cluster-cost} over $k$ clusters, 
\[
\sum_{i=1}^k z_i
=\sum\limits_{i=1}^{k}(2t-1)m_i-(t-1)-\cost(\mathcal{F}_i)\geq 
(2t-1)|E_{\rho}|-(t-1)k-\left((2t-1)M-\left(1-\frac{1}{8t}\right)kd\right).
\]
Using~\eqref{eq:most_edges_are_good} and simplifying gives the result.
\end{proof}

\subsubsection{Dominant Color in Each Cluster}
Our next goal is to show that the properties of the coloring $\rho$ imply that $\kappa_i$ is negligible. In that case,~\Cref{claim:sum_of_z} implies that on average $D_i$ is large, giving that $d_{i,v}$ is abnormally large for some $v$'s.

\begin{claim}\label{clm:kappa-small}
For every $i\in[k]$ it holds that 
$(m_i-1)\kappa_i=o(d)$.
\end{claim}

\begin{proof}
For distinct colors $a,b$, let $q_{a,b}$ denote the number of
hyperedges in $\mathcal F_i$ containing both $a$ and $b$. By the
properties of $\rho$ we have $q_{a,b} = o(d)$.
It follows that
\[
\sum_{\{e,f\}\subseteq\mathcal F_i}
\max(|\rho(e)\cap\rho(f)|-1,0)
\leq
\sum_{a<b}\binom{q_{a,b}}2
\leq
o(d)\cdot\sum_{a<b}q_{a,b}
=o\left(d\binom{t}{2}m_i\right).
\]
Dividing by $m_i/2$ gives
$(m_i-1)\kappa_i
\leq
o\left(\binom{t}{2}d\right)
=o(d)$.
\end{proof}

For every $i\in[k]$, let $c_i$ be a color appearing in the maximum
number of hyperedges of $\mathcal F_i$, and let
\[
h_i=|\{e\in\mathcal F_i:c_i\in\rho(e)\}|,
\qquad
R_i=m_i-h_i.
\]
Thus, $R_i$ is the number of hyperedges in $\mathcal F_i$ that do not
contain its dominant color. We next show that in total, the nondominant colors appear in only few hyperedges.

\begin{claim}\label{clm:few-nondominant}
We have $\sum_{i=1}^k R_i=o(M)$.
\end{claim}
\begin{proof}
Our main goal is to show that
\begin{equation}\label{eq:gamma-vs-R}
(m_i-1)\gamma_i
\geq
\frac{R_i}{4t}-240t\chi\log_2 k,
\end{equation}
and we first explain why that completes the proof.
Rearranging and summing gives 
\[
\sum\limits_{i=1}^{k} R_i
\leq O(k\chi\log_2 k) + 
4t\sum\limits (m_i-1)\gamma_i
\leq o(M) + 4t\sum\limits_{i=1}^{k} D_i-z_i + (m_i-1)\kappa_i
\leq 
o(M)+4t\sum\limits_{i=1}^{k} D_i-z_i,
\]
where we used~\Cref{clm:kappa-small} and $kd = o(M)$. 
Note that $\sum\limits_{i=1}^{k}D_i\leq \sum\limits_{i=1}^{k}\frac{d}{|\mathcal{F}_i|}\sum\limits_{v\in V}d_{i,v} = tdk$, and by~\Cref{claim:sum_of_z} we have 
$\sum\limits_{i=1}^{k}z_i\geq 0$, so the right hand side at most $o(M)+O(tdk) = o(M)$.

We now establish~\eqref{eq:gamma-vs-R}, and for that we denote by $Z_i=\binom{m_i}{2}\gamma_i$ the number of unordered pairs of hyperedges in
$\mathcal F_i$ having disjoint color sets. Fix 
$e\in\mathcal F_i$ not containing $c_i$, and note that for each of the $t$ colors
of $e$, at most $120\chi\log_2 k$ hyperedges contain both that color and $c_i$.
Thus $e$ shares a color with at most $120t\chi\log_2 k$ of the $h_i$ hyperedges
containing $c_i$, and hence 
\[
Z_i\geq R_i(h_i-120t\chi\log_2 k).
\]
If $h_i\geq \frac{m_i}{2t}$ then we immediately get~\eqref{eq:gamma-vs-R}, so we henceforth assume that 
$h_i<\frac{m_i}{2t}$. Since $c_i$ is a most frequent color, we have
$r_{i,c}\leq h_i$ for every $c$, so the number of pairs of
hyperedges whose color sets intersect is at most
$
\sum_{c\in[k]}\binom{r_{i,c}}2
\leq
\frac{h_i}{2}\sum_{c\in[k]}r_{i,c}
=
\frac{th_im_i}{2}$. Thus,
\[
Z_i
\geq
\binom{m_i}{2}-\frac{th_im_i}{2}
\geq 
\binom{m_i}{2}-\frac{m_i^2}{4}
= \frac{m_i^2-2m_i}{4}
\geq m_i\frac{R_i}{8t},
\]
and multiplying by $\frac{2}{m_i}$ implies~\eqref{eq:gamma-vs-R}.
\end{proof}

\subsubsection{Assigning Hyperedges to Vertices}
For every $e\in\mathcal F_i$ containing $c_i$, let $v(e)$ be the
unique vertex in $e$ such that $\rho(v(e))=c_i$. For every vertex $v$
with $\rho(v)=c_i$, let $a_{i,v}:=|\{e\in\mathcal F_i:v(e)=v\}|$.
We wish to show that these assignments are concentrated on relatively
few vertices. First, note that in the completeness case, a cluster consists of hyperedges assigned to vertices of a fixed color, leading to a saving of roughly $d$ (due to the coordinates corresponding to these vertices being large in $\sum\limits_{e\in F_i}p_e$), and we want to argue that something similar must happen in our case.~\Cref{claim:sum_of_z} gives a lower bound on the saving in the clusters, but in principle it could be the case that all of the savings come from a single cluster. The following claim shows that the saving $z_i$ never exceeds $d$ by much, thereby eliminating this possibility and asserting that the saving must be spread among the different clusters. 

\begin{claim}\label{clm:bounded-saving}
For every $i\in[k]$ we have $z_i\leq(1+o(1))d$.
\end{claim}
\begin{proof}
To upper bound $z_i$ we need to upper bound $D_i$, and to do that we first consider an auxiliary parameter $D_i'$. Consider only the $h_i$ hyperedges in $\mathcal F_i$ that
contain the dominant color $c_i$, let $d'_{i,v}$ denote the degree of
$v$ among these hyperedges, and define $D_i':=\frac1{m_i}\sum_{v\in V}(d'_{i,v})^2$.
Every such hyperedge contains exactly one vertex of color $c_i$.
Since $d'_{i,v}\leq d$ for every $v$ and  $\sum_{v:\rho(v)=c_i}d'_{i,v}=h_i$,
the contribution of vertices of color $c_i$ is at most $\sum_{v:\rho(v)=c_i}(d'_{i,v})^2
\leq dh_i$.
Moreover, every vertex of color different from $c_i$ has degree at
most $3\chi$ among these hyperedges. Since there are $(t-1)h_i$
incidences with such vertices, their contribution is at most $3\chi(t-1)h_i$.
Hence
\[
D_i'
\leq
\frac{dh_i+3\chi(t-1)h_i}{m_i}
\leq
d+3\chi(t-1).
\]

We now add back the remaining $R_i$ hyperedges. Adding one hyperedge
increases the degree of each of its $t$ vertices by one, and hence
increases the sum of squared degrees by at most $2td$. Therefore
\[
D_i
\leq
D_i'+\frac{2tdR_i}{m_i}
\leq
d+3\chi(t-1)+\frac{2tdR_i}{m_i}.
\]
If $m_i\geq 8t^2d$, then we get that
\[
z_i\leq D_i -(m_i-1)\gamma_i+(m_i-1)\kappa_i
\leq d+3\chi(t-1)+\frac{R_i}{4t}-(m_i-1)\gamma_i+o(d)
\leq (1+o(1))d,
\]
where we used~\eqref{eq:gamma-vs-R} to bound $\frac{R_i}{4t}-(m_i-1)\gamma_i$ and then~\Cref{clm:kappa-small}.
If $m_i<8t^2d$, applying~\Cref{lem:hypergraph-second-moment} with $\mathcal{F}_i$ gives that 
$D_i\leq d+O_t(\sqrt{\chi m_i}) = (1+o(1))d$, and again we get $z_i\leq D_i+(m_i-1)\kappa_i\leq (1+o(1))d$ using~\Cref{clm:kappa-small}.
\end{proof}

We already know that the total saving $\sum\limits_{i=1}^{k}z_i$ is large by~\Cref{claim:sum_of_z}, and the next claim shows that it is still large when we weight the clusters according to the popularity of the dominant color it them.
\begin{claim}\label{clm:weighted-saving}
We have $\sum_{i=1}^k h_i z_i
\geq
\left(\frac78-o(1)\right)Md$.
\end{claim}
\begin{proof}
By the balance of the color classes $h_i\leq d|\rho^{-1}(c_i)|
\leq(t+o(1))\frac{M}{k}$,
while~\Cref{clm:bounded-saving} gives $z_i\leq(1+o(1))d$. It follows that
\[
\sum\limits_{i=1}^{k}((1+o(1))d-z_i)h_i\leq 
(t+o(1))\frac{M}{k}\sum\limits_{i=1}^{k}(1+o(1))d-z_i
\leq 
\left(\frac{1}{8}-o(1)\right)dM,
\]
where we used~\Cref{claim:sum_of_z} in the last transition. Rearranging gives 
$\sum\limits_{i=1}^{k}h_iz_i
\geq (1+o(1))d\sum\limits_{i=1}^{k}h_i-\left(\frac{1}{8}+o(1)\right)dM$. Using~\Cref{clm:few-nondominant} to note that $\sum_i h_i
=
|E_\rho|-\sum_iR_i
=
(1-o(1))M$ finishes the proof.
\end{proof}

We now conclude that dominant vertices are not too spread.
\begin{claim}\label{clm:assignment-concentration}
We have $\sum_i\sum_{v:\rho(v)=c_i}a_{i,v}^2
\geq
\left(\frac78-o(1)\right)Md$.
\end{claim}
\begin{proof}
Consider only the $h_i$ hyperedges of $\mathcal{F}_i$ that have the color $c_i$, and let $d'_{i,v}$
be the degree of $v$ among them. Since removing one hyperedge
decreases the sum of squared degrees by at most $2td$, we have
\[
\sum_{v\in V}(d'_{i,v})^2
\geq m_iD_i-2tdR_i.
\]
For vertices of color $c_i$, we have $d'_{i,v}=a_{i,v}$.
On the other hand, as in the proof of~\Cref{clm:bounded-saving},
vertices of other colors contribute at most $3\chi(t-1)h_i$
to the sum of squared degrees. Consequently,
\[
\sum_{v:\rho(v)=c_i}a_{i,v}^2
\geq
m_iD_i-2tdR_i-3\chi(t-1)h_i.
\]
By the definition of $z_i$ and $\gamma_i\geq0$, we have
$D_i\geq z_i-(m_i-1)\kappa_i$, and since $m_i\geq h_i$ we get that
\[
\sum_{v:\rho(v)=c_i}a_{i,v}^2
\geq
h_i z_i-h_i(m_i-1)\kappa_i
-2tdR_i-3\chi(t-1)h_i.
\]
Summing over $i$, the first term is at least
$(7/8-o(1))Md$ by~\Cref{clm:weighted-saving}, while
the remaining terms are $o(Md)$ by~\Cref{clm:kappa-small},
\Cref{clm:few-nondominant}, and $\chi=o(d)$.
\end{proof}

\subsubsection{Choosing the Set $S$}
For every $v\in V$, let $A_v=\sum_{i:c_i=\rho(v)}a_{i,v}$
be the total number of hyperedges assigned to $v$. Since all $a_{i,v}$ are nonnegative, we get from~\Cref{clm:assignment-concentration} that
\begin{equation}\label{eq:A-second-moment}
\sum_{v\in V}A_v^2
\geq 
\sum\limits_{v\in V}\sum_{i:c_i=\rho(v)}a_{i,v}^2
\geq
\left(\frac78-o(1)\right)Md.
\end{equation}

Let $S$ consist of the $N/t$ vertices with the largest values of
$A_v$. The following claim completes the soundness analysis.
\begin{claim}
    $S$ covers more than $3M/4$ hyperedges.
\end{claim}
\begin{proof}
For every $v\in V$, define $x_v:=A_v/d$. Since $A_v\leq d$ and
$\sum_v A_v\leq M=Nd/t$, we have
\[
0\leq x_v\leq1,
\qquad
\sum_vx_v\leq\frac Nt.
\]
Thus $(x_v)_{v\in V}$ is a fractional choice of at most $N/t$
vertices. Since $S$ consists of the $N/t$ vertices with the largest
values of $A_v$, it captures at least as much load as this
fractional choice, so using~\eqref{eq:A-second-moment} we get
\[
\sum_{v\in S}A_v
\geq
\sum_vx_vA_v
=
\frac1d\sum_vA_v^2
\geq
\left(\frac78-o(1)\right)M
>
\frac{3M}{4}.\qedhere
\]
\end{proof}

\section{Lower Bound for Euclidean $k$-median under Almost ETH}
\label{sec:euclidean-kmedian-lower-bound}
In this section we prove~\Cref{thm:k_median}, restated below.

\begin{theorem}
\label{thm:bmv-kmedian-formal}
Assuming Almost ETH, the following holds for every constant $\beta\in(0,1)$. Let $\tilde{\eps}\colon \mathbb{N}\to (0,1)$, $\tilde{k}\colon \mathbb{N}\to\mathbb{N}$ be polynomial time computable functions of the input size satisfying:
\begin{enumerate}
    \item $\frac{\tilde{k}(n)}{\tilde{\eps}(n)} = o\left(\frac{n}{\log^L n}\right)$ and the sequence $\frac{\tilde{k}(n)}{\tilde{\eps}(n)}$ has bounded consecutive ratios.
    \item $\omega(\log n)\leq \frac{1}{\tilde{\eps}(n)}\leq o\left(\frac{1}{(\log n)^{2L} }\right)\tilde{k}(n)$.
\end{enumerate}
Then for sufficiently large $n$, setting $\eps = \tilde{\eps}(n)$ and $k = \tilde{k}(n)$, Euclidean
$k$-median on $n$ points in $\mathbb R^d$ has no
$(1+\varepsilon)$-approximation running in time    $2^{(k/\varepsilon)^{1-\beta}}
    \operatorname{poly}(n,d)$.
\end{theorem}

In the proof of~\Cref{thm:bmv-kmedian-formal} we will transform the $k$-means lower bound from~\Cref{thm:bmv-kmeans-formal}.
The main idea is to give each point a new, unique coordinate of a large magnitude which we denote by $T$ (a similar idea was used in~\cite{Shenmaier21}). For a center that uses none of these new coordinates, a point whose original squared distance to the center is $x$ now has distance 
\[
\sqrt{T^2+x} = T\sqrt{1+\frac{x}{T^2}} = T\left(1+\frac{x}{2T^2} + O\left(\frac{x}{T^2}\right)^2\right)
= T + \frac{1}{2T}x + O\left(\frac{x^2}{T^3}\right)
\] 
to its center. Thus, if $x$ significantly smaller than $T$, we get that the distance of the point from the center relates linearly to $x$. This suggest that as long as the centers do not use the new coordinates, one can directly relate the $k$-median value of the new instance with the $k$-mean value of the original instance. This turns out to be true but requires a much more delicate analysis due to the accumulation of errors in the Taylor expansion above across all of points. To finish the proof, one has to argue that the centers cannot benefit much from using the private coordinates, which holds because each point gets its own private coordinate.

\vspace{-1ex}
\paragraph{Construction.}
Fix a $t$-uniform $d$-regular hypergraph
$\HH=(V,E)$ on $N$ vertices as in
\Cref{thm:hypergraph-to-kmeans}, and fix the family of colorings $\mathcal{C}_{N,k}$  from~\Cref{lem:enumerable-good-colorings}. We drop from $\mathcal{C}_{N,k}$ any coloring failing one of the first four items (this can be done as each one of them can be checked in in polynomial time), and by~\Cref{lem:enumerable-good-colorings} we still have $\mathcal{C}_{N,k}\neq\emptyset$.
For each retained coloring $\rho\in \mathcal{C}_{N,k}$, let
$p_1,\ldots,p_m$,
be the point from $P_{\rho}$ from \Cref{sec:hypergraph-kmeans-construction}. For an integer $T>0$ (to be determined later), define the new point set $Q_\rho:=\{q_1,\ldots,q_m\}\subseteq \mathbb R^{N+k+m}$ as follows:
\begin{equation*}
q_j:=(p_j,T\mathbf e_j)\in\mathbb R^{N+k+m} \quad \forall j\in [m].
\end{equation*}
Thus each point has $2t+1$ nonzero coordinates and we refer to the last $m$ coordinates as private coordinates. This completes the description of the reduction, and the heart of the analysis of the reduction is in the following lemma.

\begin{lemma}
\label{lem:kmedian-objective-comparison}
For every $t\geq2$ and $\eta>0$ there is an integer
$T=T(t,\eta)>0$ such that for
every coloring $\rho\in\mathcal{C}_{N,k}$, we have:
\[
\left|
\OPT^{1}_k(Q_\rho)
-
\frac{1}{2r}\OPT^{2}_k(P_\rho)
-
\frac{2r^2 - (2t-1)}{2r}M
\right|
\leq \frac{kd}{2r}\eta,
\]
where $M = \frac{Nd}{t}$ and $r=\sqrt{T^2+2t-1}$.
\end{lemma}

\subsection{Proof of~\Cref{thm:bmv-kmedian-formal} Assuming~\Cref{lem:kmedian-objective-comparison}}

Suppose towards contradiction that such an approximation algorithm
exists. Take a $t$-uniform hypergraph $\HH$ from~\Cref{thm:bmv-hypergraph-source} with error parameter $\frac{1}{100}$, $N$ vertices and degree $d$. 
As before, we take the first $q$ such that $\frac{\widetilde{k}(q)}{\widetilde{\eps}(q)}\geq BtN$, where $B$ is a sufficiently large constant to be determined. Using the same argument as in~\Cref{thm:bmv-kmeans-formal} we have that $q\geq N$, $\frac{\widetilde{k}(q)}{\widetilde{\eps}(q)} = \Theta(N)$ and $M=o(q)$, and we denote for notational simplicity $k = \tilde{k}(q)$, $\eps = \widetilde{\eps}(q)$ and $k'=k-1$. 
We take the collection of colorings $\mathcal{C}_{N,k'}$, from~\Cref{lem:enumerable-good-colorings}, and we only keep colorings satisfying the first four items.

Choose $\eta=1/(64t)$ and apply~\Cref{lem:kmedian-objective-comparison} to get $T$, $r=\sqrt{T^2+2t-1}$ for the collections $(Q_\rho)_{\rho\in\mathcal{C}_{N,k'}}$ above. Recalling that $|Q_{\rho}|\leq M = o(q)$, we pad each $Q_{\rho}$ by setting
$W=\lfloor(q-1)/|Q_{\rho}|\rfloor$ and forming $Q_{\rho}'$ by including $W$ copies of each point from $Q_{\rho}$ where we append a new $0$ coordinate, and including $q-W|Q_{\rho}|$ copies of the  point
$x=Hq\mathbf e_{\mathrm{new}}$, where
$H>2\sqrt{T^2+2t}$ is a fixed integer. As in the proof of~\Cref{thm:bmv-kmeans-formal} we get that
\[
\OPT_k^{1}(Q'_\rho)
=
W\OPT_{k'}^{1}(Q_\rho),
\]
and we now use the guarantee of~\Cref{lem:kmedian-objective-comparison} to calculate the gap between the YES and the NO instances.
In the YES case, by~\Cref{thm:hypergraph-to-kmeans},
there is some $\rho$ such that $\OPT_{k'}^{2}(P_\rho)
\leq (2t-1)M-(1-o(1))k'd$, so by~\Cref{lem:kmedian-objective-comparison}
\begin{align*}
\OPT_{k'}^{1}(Q_\rho)
\leq
\left(r-\frac{2t-1}{2r}\right)M
+\frac{\OPT_{k'}^{2}(P_\rho)}{2r}
+\frac{\eta k'd}{2r}
&\leq rM-\left(1-\eta-o(1)\right)\frac{k'd}{2r}\\
&\leq rM-\left(1-\frac1{32t}\right)\frac{k'd}{2r},
\end{align*}
which we denote by $U$. In the last transition we used the fact that $N$ is sufficiently large. 
In the NO case, by~\Cref{thm:hypergraph-to-kmeans},
for every $\rho\in\mathcal{C}_{N,k'}$ we have $\OPT_{k'}^{2}(P_\rho)
\geq (2t-1)M-\left(1-\frac1{8t}\right)k'd$, so by~\Cref{lem:kmedian-objective-comparison}
\begin{align*}
\OPT_{k'}^{1}(Q_\rho)
\geq
\left(r-\frac{2t-1}{2r}\right)M
+\frac{\OPT_{k'}^{2}(P_\rho)}{2r}
-\frac{\eta k'd}{2r}
&\geq
rM-\left(1-\frac1{8t}+\eta\right)\frac{k'd}{2r}\\
&=
rM-\left(1-\frac7{64t}\right)\frac{k'd}{2r},
\end{align*}
which we denote by $L$. It follows that the ratio between the two cases after padding is also $L/U$, and we now calculate it. Note that $L-U=\frac{5k'd}{128tr}>\frac{k'd}{32tr}$, so
\[
\frac{L}{U}\geq 1+\frac{k'd}{32trU}
=1+\Theta\left(\frac{BN\eps d}{r^2M}\right)
=1+\Theta\left(\frac{Bt}{r^2}\right)\eps
\geq 1+\eps
\]
for sufficiently large $B$.

\subsection{Proof of~\Cref{lem:kmedian-objective-comparison}}
Fix a coloring $\rho\in\mathcal{C}_{N,k}$ and denote
$m=|P_\rho|=|Q_\rho|$. We prove the upper bound and the lower bound separately.

\subsubsection{Proof of the Upper bound}
Consider an optimal $k$-means clustering of $P_\rho$ with centers $s_1,\ldots,s_k$. Let $s_1',\ldots,s_k'$ be the points $s_1,\ldots,s_k$ where we append $0$'s on all of the new coordinates, and consider it as centers for $Q_\rho$. Since the second order derivatives of $x\mapsto\sqrt{T^2+x}$ are negative, we get by Taylor approximation around $2t-1$ that
\[
\sqrt{T^2+x}
\leq r+\frac{1}{2r}(x-(2t-1)).
\]
Applying this inequality to each point in $Q_{\rho}$ and then summing gives
\[
\OPT_k^{1}(Q_\rho)
\leq rm+\frac{\OPT_k^{2}(P_\rho)-(2t-1)m}{2r}
\leq
\left(r-\frac{2t-1}{2r}\right)M
+\frac{\OPT_k^{2}(P_\rho)}{2r}.
\]
where we used $m\leq M$.

\subsubsection{Proof of the Lower bound} 
Fix an optimal $k$-median clustering of $Q_\rho$, and let
$\mathcal F_1,\ldots,\mathcal F_k$ be the corresponding clusters
of hyperedges. We will use the following notations from~\Cref{sec:hypergraph-kmeans-soundness}:  let $m_i = |\mathcal{F}_i|$, let $c_i$ be the most popular color in $\mathcal{F}_i$, let $R_i$ be the number of hyperedges in $\mathcal{F}_i$ that do not contain the color $c_i$ and let $\mu_i = \frac{1}{m_i}\sum\limits_{e\in\mathcal{F}_i}p_e$. 
The following lemma breaks the cost of each cluster into its $k$-means cost, the effect of the private coordinates, and the terms arising from the Taylor approximation and the choice of center.
\begin{lemma}\label{lem:kmedian-cost-decomposition}
For every center $(y,b)$, we have
\begin{align*}
\sum_{e\in\mathcal{F}_i}\|q_e-(y,b)\|_2
=&
\underbrace{
\left(r-\frac{2t-1}{2r}\right)m_i
+\frac{\cost(\mathcal F_i)}{2r}
}_{\text{main term}}
+
\underbrace{\frac{m_i}{2r}\|y-\mu_i\|_2^2}
_{\text{distance from $\mu_i$}}-
\underbrace{
\frac1{2r}\sum_{e\in \mathcal{F}_i}
\left(\sqrt{T^2+\|p_e-y\|_2^2}-r\right)^2
}_{\text{Taylor error}}\\
&-
\underbrace{
\sum_{e\in \mathcal{F}_i}\left(
\sqrt{T^2+\|p_e-y\|_2^2}
-\|q_e-(y,b)\|_2
\right)
}_{\text{effect of private coordinates}}.
\end{align*}
\end{lemma}
\begin{proof}
Our goal is to relate the $k$-median cost of the cluster to its
$k$-means cost. We first consider a center $(y,0)$ that does not
use the private coordinates. In this case, the distance of
$q_e$ from the center is
\[
\|q_e-(y,0)\|_2=\sqrt{T^2+\|p_e-y\|_2^2}.
\]
Its
first-order Taylor approximation around $2t-1$ is $r+\frac{\|p_e-y\|_2^2-(2t-1)}{2r}$, and we want to calculate the difference between them explicitly. Recalling that
$r^2=T^2+2t-1$, we have for every $x\geq0$
\[
r+\frac{x-(2t-1)}{2r}-\sqrt{T^2+x}
=
\frac{(\sqrt{T^2+x}-r)^2}{2r}.
\]
Thus, summing over the cluster gives
\[
\sum_{e\in\mathcal{F}_i}\|q_e-(y,0)\|_2
={}
\left(r-\frac{2t-1}{2r}\right)m_i
+\frac1{2r}\sum_{e\in\mathcal{F}_i}\|p_e-y\|_2^2-\frac1{2r}\sum_{e\in \mathcal{F}_i}
\left(\sqrt{T^2+\|p_e-y\|_2^2}-r\right)^2.
\]

The first sum of squared distances is almost the $k$-means
cost of the cluster, except that $y$ need not be its center.
By~\Cref{lem:folklore}, we have
\[
\sum_{e\in \mathcal{F}_i}\|p_e-y\|_2^2
=
\cost(\mathcal F_i)+m_i\|y-\mu_i\|_2^2,
\]
and plugging this above we get
\begin{align*}
\sum_{e\in\mathcal{F}_i}\|q_e-(y,0)\|_2
&=
\left(r-\frac{2t-1}{2r}\right)m_i
+
\frac1{2r}\cost(\mathcal F_i)+\frac{m_i}{2r}\|y-\mu_i\|_2^2\\
&-\frac1{2r}\sum_{e\in \mathcal{F}_i}
\left(\sqrt{T^2+\|p_e-y\|_2^2}-r\right)^2.
\end{align*}
Finally, we account for the fact that the centers of $\mathcal{F}_i$ may use the private coordinates. For an arbitrary center $(y,b)$, we have
\[
\sum_{e\in \mathcal{F}_i}\|q_e-(y,b)\|_2
=
\sum_{e\in \mathcal{F}_i}\|q_e-(y,0)\|_2
-\sum_{e\in \mathcal{F}_i}
\left(\sqrt{T^2+\|p_e-y\|_2^2}
-\|q_e-(y,b)\|_2\right).
\]
Combining this with the previous calculation gives exactly
the claimed decomposition.
\end{proof}
We now bound the error terms in~\Cref{lem:kmedian-cost-decomposition}. We start with the contribution of the private coordinates.
\begin{claim}\label{clm:kmedian-private}
For an
optimal center $(y,b)$ of $\mathcal F_i$, we have
\[
\sum_{e\in\mathcal F_i}
\left(
\sqrt{T^2+\|p_e-y\|_2^2}
-\|q_e-(y,b)\|_2
\right)\leq 2T.
\]
\end{claim}
\begin{proof}
As an optimal $1$-median center can be chosen
in the convex hull of the points in its cluster, we assume that $(y,b)$ is in the convex hull of $\mathcal{F}_i$, and the sum of all coordinates of $b$ are non-negative and they sum up to $T$. For every $e\in\mathcal F_i$, we have $\|q_e-(y,0)\|_2^2-\|q_e-(y,b)\|_2^2
=2Tb_e-\|b\|_2^2\leq2Tb_e$, and as $\|q_e-(y,0)\|_2\geq T$, we get
\[
\|q_e-(y,0)\|_2-\|q_e-(y,b)\|_2
\leq2b_e.
\]
Summing over the cluster gives the claim.
\end{proof}
We next bound the Taylor error term on~\Cref{lem:kmedian-cost-decomposition}, and we show that its negative contribution is almost entirely offset by the distance to $\mu_i$.
\begin{claim}\label{clm:kmedian-taylor-error}
For sufficiently large $T = T(t,\eta)$, and every $y$ in the convex hull
of the points of $\mathcal F_i$, we have
\[
\frac{1}{2r}\sum_{e\in\mathcal F_i}
\left(\sqrt{T^2+\|p_e-y\|_2^2}-r\right)^2
\leq
\frac{m_i}{2r}\|y-\mu_i\|_2^2
+O_t\left(\frac{R_i+d}{T^3}\right).
\]
\end{claim}
\begin{proof}
Since $r^2=T^2+2t-1$, we have
\begin{equation}\label{eq:taylor}
\frac1{2r}\left(\sqrt{T^2+x}-r\right)^2
=
\frac{(x-(2t-1))^2}
{2r(\sqrt{T^2+x}+r)^2}\leq \frac{(x-(2t-1))^2}{8T^3}.
\end{equation}
Thinking of $x=\|p_e-y\|_2^2$, our goal is to show that $\|p_e-y\|_2^2$ are typically close to $2t-1$.
Towards this end, let $u_i=e_{N+c_i}$ be the unit vector corresponding
to the dominant color $c_i$, and note that
$\|p_e-u_i\|_2^2$ is $2t-1$ if $e$ contains $c_i$,
and is $2t+1$ otherwise. In particular, only $R_i$
hyperedges have squared distance different from $2t-1$.
We get
\begin{align*}
\|p_e-y\|_2^2-(2t-1)
&=\|p_e-u_i\|_2^2-(2t-1)-2\langle p_e-u_i,y-u_i\rangle
+\|y-u_i\|_2^2
\\
&=2\cdot 1_{c_{i}\not\in\rho(e)}-2\langle p_e-u_i,y-u_i\rangle
+\|y-u_i\|_2^2.
\end{align*}
Using Cauchy-Schwarz and the fact that $\|p_e-u_i\|_2,\|y-u_i\|_2\leq \sqrt{2t+1}$ gives 
\[
|\|p_e-y\|_2^2-(2t-1)|
\leq 2\cdot 1_{c_{i}\not\in\rho(e)}+3\sqrt{2t+1}\|y-u_i\|_2.
\]
Squaring, using $(a+b)^2\leq 2a^2+2b^2$ and summing gives
\[
\sum_{e\in\mathcal F_i}
\left(\|p_e-y\|_2^2-(2t-1)\right)^2
\leq 8 R_i+ 18(2t+1)m_i\|y-u_i\|_2^2.
\]
We next upper bound $\|y-u_i\|_2^2$. Using the triangle inequality we get
\[
m_i\|y-u_i\|_2^2
\leq
2m_i\|y-\mu_i\|_2^2
+
2m_i\|\mu_i-u_i\|_2^2.
\]
By~\Cref{lem:folklore} and~\Cref{eq:soundness-cluster-cost} we get
\[
m_i\|\mu_i-u_i\|_2^2=\sum\limits_{e\in \mathcal{F}_i}\|p_e-u_i\|_2^2-\cost(\mathcal{F}_i)
= (2t-1)m_i+2R_i-\cost(\mathcal{F}_i)
= 2R_i+
(t-1)+z_i,
\]
which is at most $2R_i+O_t(d)$ using~\Cref{clm:kappa-small} and $D_i\leq td$. Plugging this above and then propagating further gives
\[
\sum_{e\in\mathcal F_i}
\left(\|p_e-y\|_2^2-(2t-1)\right)^2
\leq O_t(R_i)+ 18(2t+1)(2m_i\|y-\mu_i\|_2^2+O_t(d)).
\]
Plugging $x=\|p_e-y\|_2^2$ in~\eqref{eq:taylor} and using the above to upper bound the right hand side of~\eqref{eq:taylor}, we get that the left hand side of the claim is upper bounded by
\[
O_{t}\left(\frac{R_i+d}{T^3}\right)+\frac{9(2t+1)}{2T^3}m_i\|y-\mu_i\|_2^2
\leq O_{t}\left(\frac{R_i+d}{T^3}\right)+\frac{m_i}{2r}\|y-\mu_i\|_2^2,
\]
where the last inequality is for large enough $T$.
\end{proof}

Plugging~\Cref{clm:kmedian-private} and~\Cref{clm:kmedian-taylor-error} into~\Cref{lem:kmedian-cost-decomposition} and summing over $i$ gives the bound
\begin{equation}\label{eq:finish_proof_median}
    \OPT^{1}_{k}(Q_{\rho})
    \geq \left(r-\frac{2t-1}{2r}\right)\sum\limits_{i=1}^{k}m_i
+\frac{1}{2r} \sum\limits_{i=1}^{k}\cost(\mathcal{F}_i)
- \frac{1}{T^3}O_{t}\left(dk+\sum\limits_{i=1}^{k}R_i\right)
-2Tk,
\end{equation}
and we next bound $\sum\limits_{i=1}^{k}R_i$. Rearranging~\eqref{eq:gamma-vs-R} we get that
\[
\sum\limits_{i=1}^{k}R_i\leq 
4t(240tk\chi\log_2 k + \sum\limits_{i=1}^{k}(m_i-1)\gamma_i)
= O_{t}(kd) + 4t\sum\limits_{i=1}^{k} D_i-z_i+(m_i-1)\kappa_i
\leq 
O_{t}(kd)-4t\sum\limits_{i=1}^{k}z_i
\]
where we used~\Cref{clm:kappa-small} and $D_{i}\leq td$ in the last transition. Using $z_i =(2t-1)m_i -(t-1)-\cost(\mathcal{F}_i)$ above and plugging that into~\eqref{eq:finish_proof_median} gives
\begin{align*}
\OPT^{1}_k(Q_{\rho})
&\geq 
r\sum\limits_{i=1}^{k}m_i
+\left(\frac{1}{2r}-\frac{O_{t}(1)}{T^3}\right) \left(\sum\limits_{i=1}^{k}\cost(\mathcal{F}_i)-(2t-1)m_i\right)
-\frac{1}{T^3}O_{t}(kd)-2Tk\\
&\geq r\sum\limits_{i=1}^{k}m_i
+\left(\frac{1}{2r}-\frac{O_{t}(1)}{T^3}\right) \left(\OPT^{2}_k(P_{\rho})-(2t-1)\sum\limits_{i=1}^{k}m_i\right)
-\frac{1}{T^3}O_{t}(kd)-2Tk,
\end{align*}
where we used $\frac{1}{2r}-\frac{O_{t}(1)}{T^3}\geq 0$ as $T$ is sufficiently large and 
$\sum\limits_{i=1}^{k}\cost(\mathcal{F}_i)\geq \OPT^{2}_k(P_{\rho})$. Considering the $k$-clustering corresponding to the $k$-colors and choose as centers the vectors $e_{N+c}$ for $c\in [k]$ we get $\OPT^{2}(P_{\rho})-(2t-1)\sum\limits_{i=1}^{k}m_i\leq 0$, giving us
\[
\OPT^{1}_k(Q_{\rho})
\geq 
r\sum\limits_{i=1}^{k}m_i
+\frac{1}{2r}\left(\OPT^{2}_k(P_{\rho})-(2t-1) \sum\limits_{i=1}^{k}m_i\right)
-\frac{1}{T^3}O_{t}(kd)-2Tk.
\]
By~\eqref{eq:most_edges_are_good} we have that $\sum\limits_{i=1}^{k}m_i\geq M - O_{t}\left(\frac{M}{k}\right)\geq M-o(kd) \geq M-\frac{kd}{rT^3}$
and rearranging gives that
\[
\OPT^{1}_{k}(Q_{\rho})
-\left(r-\frac{2t-1}{2r}\right)M
-\frac{1}{2r}\OPT^{2}_k(P_{\rho})
\geq 
-\frac{1}{T^3}O_{t}(kd)-2Tk
\geq -\frac{dk}{2r}\eta.
\]

\bibliographystyle{alpha}
\bibliography{refs}
\appendix
\section{Disproof of XXH}\label{sec:XXH_false}
In this section we show that the XXH assumption of~\cite{CKSS26} is false.
For a graph $G=(V,E)$ and $W\subseteq V$, let $N_G(W)$ be the set of
vertices adjacent to at least one vertex in $W$.

\begin{definition}
\label{def:smallLetExpander}
Let $\alpha>0$, let $G=(V,E)$ be a $d$-regular graph on $n$ vertices,
and let $K$ be a positive integer. We say that $G$ is a
$(K,\alpha)$-\emph{small-set vertex expander} if every $W\subseteq V$ of size at most $n/K$ satisfies
$|N_G(W)|\geq(1-\alpha)d|W|$.
\end{definition}

The hypothesis of~\cite{CKSS26}, parameterized by $\delta,\zeta,\alpha,B$ and $L$, can be stated as follows.

\begin{hypothesis}
\label{hyp:xxh}
Fix $\delta,\zeta,\alpha\in(0,1)$, $B>0$, $L>1$. For all $\beta\in(0,1)$, there is no randomized $2^{n^{1-\beta}}$-time algorithm that given a graph $G=(V,E)$, distinguishes
the following cases with probability at least $0.9$:
\begin{enumerate}
\item \textbf{Completeness:} There is a set $S\subseteq V$ of size $n/2$
that covers at least $(1-\zeta)|E|$ edges.
\item \textbf{Soundness:} Every set $S\subseteq V$ of size $n/2$ leaves
at least $\delta|E|$ edges uncovered.
\end{enumerate}
In both cases, $G$ is a simple $\log^{L} n$-regular graph
and a $(\log^{B}n,\alpha)$-small-set vertex expander.
\end{hypothesis}

The reduction in~\cite{CKSS26} uses~\Cref{hyp:xxh} with
$\zeta\leq\delta^5/10<\delta^2/2$, and we show that for such parameters the hypothesis is false. 
The key observation
is that if $S\subseteq V$ has size $n/2$, then by regularity of the graph the number of edges fully contained in $S$ and $V\setminus S$ is the same. Thus, if $S$ touches almost all edges of $G$, then $G$ has a large cut, and we can apply the Goemans-Williamson algorithm~\cite{GoemansW95} to find a large cut. 

\begin{proposition}
\label{prop:xxh-max-bisection}
If $\zeta<\delta^2/2$ and $\delta$ is sufficiently small, then there is a randomized polynomial-time
algorithm distinguishing the completeness and soundness cases
in~\Cref{hyp:xxh} with probability at least $0.9$.
\end{proposition}

\begin{proof}
Let $m:=|E|$. For every $A\subseteq V$, let $E(A)$ denote the edges with
both endpoints in $A$. If $|A|=n/2$, the regularity of the graph gives
$|E(A)|=|E(V\setminus A)|$: both sides have total sum of degrees to be  $dn/2$ and
the same number of edges crossing the cut. Thus, in the completeness
case, the promised set $S$ gives a cut with at most $2\zeta m$ edges
not crossing it.

Applying the randomized algorithm from~\cite[Section 3.1]{GoemansW95} polynomially many times, with probability at least $0.99$ we find a cut $S\subseteq V$ such that at least $(1-\sqrt{2\zeta})m>(1-\delta)m$ cross it. We may assume that $|S|\leq n/2$ (otherwise we take $\overline{S}$), and then add to it vertices arbitrarily until it has size $n/2$.
\end{proof}

\begin{remark}
\label{rem:weakXXH}
One may consider a weaker version of XXH, which is sufficient for the analysis~\cite{CKSS26}. More precisely, one could replace the set size being $n/2$ with being $(1/2+r^3)n$ for a small $r>0$, require perfect coverage in the completeness and missing at least $r^2|E|$ edges in the soundness.
However, even this weaker hypothesis is false. Indeed, for small enough $\alpha$,~\cite[Theorem 2.4]{Steurer10dgames} asserts that $G$ has at most $\tilde{O}(\sqrt n)$ eigenvalues at least $2/5$, and in the completeness case~\cite[Corollary 7.21]{BHSV25} shows how to find an independent set of size at
least $(1/2-O(r^3))n$  in time
$2^{n^{1/2+o(1)}}$. Taking the complement of this set, and adding/removing $O(r^3 n)$ vertices from it gives a set touching all but $O(r^3)\leq r^2$ fraction of edges.
\end{remark}

\end{document}